\documentclass{article}
\usepackage[utf8]{inputenc}
\usepackage[T1]{fontenc} 
\usepackage{arxiv}
\usepackage{amsmath}
\usepackage{braket}
\usepackage{amsfonts}
\usepackage{amssymb}
\usepackage{graphicx}
\usepackage{amsthm}
\usepackage{listings}
\usepackage{float}
\usepackage{caption}
\usepackage{subcaption}
\usepackage{color}
\usepackage[colorlinks=true,allcolors=blue]{hyperref}
\usepackage{chngcntr}
\usepackage{cite}
\usepackage{authblk}
\usepackage{tikz}
\usepackage{booktabs} 
\DeclareFontFamily{OMS}{oasy}{\skewchar\font48 }
\DeclareFontShape{OMS}{oasy}{m}{n}{%
	<-5.5> oasy5     <5.5-6.5> oasy6
	<6.5-7.5> oasy7     <7.5-8.5> oasy8
	<8.5-9.5> oasy9     <9.5->  oasy10
}{}
\DeclareFontShape{OMS}{oasy}{b}{n}{%
	<-6> oabsy5
	<6-8> oabsy7
	<8->  oabsy10
}{}
\DeclareSymbolFont{oasy}{OMS}{oasy}{m}{n}
\SetSymbolFont{oasy}{bold}{OMS}{oasy}{b}{n}

\DeclareMathSymbol{\smallleftarrow}     {\mathrel}{oasy}{"20}
\DeclareMathSymbol{\smallrightarrow}    {\mathrel}{oasy}{"21}
\DeclareMathSymbol{\smallleftrightarrow}{\mathrel}{oasy}{"24}

\definecolor{greennew}{rgb}{0.1, 0.3, 0.5}
\definecolor{rednew}{rgb}{0.8, 0, 0}
\definecolor{bluenew}{rgb}{0, 0, 0.8}
\definecolor{newgreen}{rgb}{0.07, 0.53, 0.37}
\DeclareRobustCommand\full{\tikz[baseline=-0.6ex]\draw[greennew,thick] (0,0)--(0.5,0);}
\DeclareRobustCommand\dashed{\tikz[baseline=-0.6ex]\draw[rednew,thick,densely dashed] (0,0)--(0.54,0);}
\DeclareRobustCommand\dotted{\tikz[baseline=-0.6ex]\draw[bluenew,thick,dashed] (0,0)--(0.54,0);}
\DeclareRobustCommand\chain {\tikz[baseline=-0.6ex]\draw[newgreen,thick,dashdotted] (0,0)--(0.54,0);}

\newcommand{\be}{\begin{equation}}
\newcommand{\ee}{\end{equation}}
\newcommand{\bea}{\begin{align}}
\newcommand{\eea}{\end{align}}
\newcommand{\ba}{\begin{array}}
\newcommand{\ea}{\end{array}}
\newcommand{\nn}{\nonumber}

\counterwithout{figure}{section}

\title{Ritus functions for graphene-like  systems with magnetic fields generated by first-order intertwining operators}

\author{Y Concha-S\'anchez\textsuperscript{1}, E D\'iaz-Bautista\textsuperscript{2}, A Raya\textsuperscript{3,4}\\
\textsuperscript{1} Facultad de Ingenier\'{\i}a Civil, Universidad Michoacana de San Nicol\'as de Hidalgo, Edificio C, Ciudad Universitaria. Francisco J. M\'ujica
	s/n. Col. Fel\'{\i}citas del R\'{\i}o. 58030, Morelia, Michoac\'an, M\'exico
\textsuperscript{2} Instituto Polit\'ecnico Nacional, UPIIH, Ciudad del Conocimiento y la Cultura, 42162 Hidalgo, Mexico\\
\textsuperscript{3} Instituto de F\'{\i}sica y Matem\'aticas, Universidad Michoacana de San Nicol\'as de Hidalgo, Edificio C-3, Ciudad Universitaria. Francisco J. M\'ujica s/n. Col. Fel\'{\i}citas del R\'{\i}o. 58040 Morelia, Michoac\'an, M\'exico\\
\textsuperscript{4} Centro de Ciencias Exactas, Universidad del Bío-Bío. Avda. Andrés Bello 720, Casilla 447, 3800708, Chillán, Chile \\
e-mail: yajaira.concha@umich.mx, ediazba@ipn.mx, alfredo.raya@umich.mx}
\date{ }

\begin{document}

\maketitle
\begin{abstract}
In this work, we construct the exact propagator for Dirac fermions in graphene-like systems immersed in external static magnetic fields with non-trivial spatial dependence. Such field profiles are generated within a first-order supersymmetric framework departing from much simpler (seed) magnetic field examples. The propagator is spanned on the basis of the Ritus eigenfunctions, corresponding to the Dirac fermion asymptotic states in the non-trivial magnetic field background which nevertheless admits a simple diagonal form in momentum space. This strategy enlarges the number of magnetic field profiles in which the fermion propagator can be expressed in a closed-form. Electric charge and current densities are found directly from the corresponding propagator and compared against similar findings derived from other methods. 
\end{abstract}

\section{Introduction}\label{sec1}
Physics of pseudo-relativistic Dirac fermions in two spatial dimensions continues to attract the attention of a vast community around the globe that considers these entities as fundamental in importance as the building blocks of the universe~\cite{MIRANSKY20151}. From the seminal work of Wallace~\cite{PhysRev.71.622}, the interest on this kind of excitations in condensed matter realms (see, for instance, ~\cite{marino_2017,shen2017topological} and references therein) has been put forward in quantum Hall~\cite{ Zhang2005, Williams2007, Novoselov2007}, high-T$_c$ superconductivity~\cite{Oliva-Leyva2017} and other bidimensional systems~\cite{Novoselov2004, Naumis2017}. In recent years, graphene~\cite{Zhang2005,Novoselov2005} and the plethora of new 2D materials~(see Refs.~\cite{doi:10.1146/annurev-matsci-070214-021034,AKINWANDE201742,BAZYLEWSKI2019287,ChengChang:2108017} for recent reviews) have increased the interest in these systems not only because of the potential technological applications, but also because of the fundamental physics that can be explored in a condensed matter physics environment~\cite{MIRANSKY20151,marino_2017,shen2017topological}. The dynamics of  pseudo-relativistic quasi-particle states has been explored under the influence of different external agents  like under strain, curvature effects and in the presence of (external or induced) electric and magnetic fields~\cite{MIRANSKY20151,Naumis2017,Roy_2012,VOZMEDIANO2010109,LIN2018161,NaumisRMF}. 

For the dynamics of Dirac fermions influenced by external electromagnetic fields  a lot of attention has been paid to understand the electronic states in background fields configurations related to uniform magnetic field, crossed electric and magnetic fields, parallel electric and magnetic fields and the plane wave electromagnetic field cases. Further configurations of static magnetic fields with spatially varying profile have also been considered from the supersymmetric quantum mechanical structure of the Dirac equation in fields of this type~\cite{Raya2010,Murguia2010}. Examples include the uniform magnetic field case (and variations including an electric field), the Scarf potential (both hyperbolic and trigonometric), and the Morse potential along one spatial dimension~\cite{Roy_2012,Concha}.
Being more precise, supersymmetry in quantum mechanics is a theoretical framework that allows to map the solutions from a stationary Schr\"odinger problem in a static one-dimensional potential to another stationary Schr\"odinger problem with a different potential that is called the supersymmetric partner of the former. Supersymmetry is realized in different manners, such as the factorization method~\cite{Mielnik1984,Berger2010} and the Darboux transformation~\cite{darboux1882,darboux1991}, which are equivalent. An interesting variant of the supersymmetric framework was  developed in~\cite{Midya2014,Celeita2020} in which rather that starting from the solutions to a Ricatti equation,  new potentials are  generated departing from the solutions of an initial wave equation.

In many physical situations, nevertheless, it is equally useful to know the corresponding propagator for these electronic states. However, because the  asymptotic states  do not correspond to plane waves,  the representation of the two-point function is cumbersome 
rendering almost impossible to write the propagator in a closed form except for a handful of examples related to the uniform electric/magnetic field either parallel or perpendicular and plane wave electromagnetic field. Alternative representations have been developed for this purpose. Among several others,  the Schwinger method~\cite{PhysRev.82.664}, the spectral representation~\cite{Suzuky05} and the Ritus method~\cite{RITUS1972,Ritus1974,Ritus1978} allow to write a closed form of the  propagator.

In this article, we revisit the construction the propagator of 2D Dirac fermions in a background static magnetic field, which is relevant to monolayer  graphene and related systems. For this purpose, we expand the propagator in the basis of Ritus functions, namely, the eigenfunctions of the operator $(\gamma\cdot\Pi)^2$ where $\Pi_\mu=p_\mu +eA_\mu$ is the canonical momentum operator that includes the effect of the external magnetic field through minimal coupling (with  $A_\mu$ denoting the corresponding vector potential and $e$ is the elementary charge) and $\gamma^\mu$ denote the $2\times 2$ covariant Dirac matrices. We consider non-trivial magnetic background fields derived within a generalization of the first order intertwining formalism of Refs.~\cite{Midya2014,Celeita2020} in which, starting from seed solutions corresponding to the Ritus eigenfunctions for the  uniform and an exponentially decaying magnetic fields~\cite{Raya2010,Murguia2010,Roy_2012,Concha}, we construct the new Ritus eigenfuctions corresponding to more intricate magnetic field profiles written in terms of highly transcendental functions. In doing so, we extend the number of cases in which the propagator for Dirac fermions in non-trivial magnetic field backgrounds can be expressed in a closed form. To achieve that goal and aiming a self-contained presentation of our findings, we have organized the remaining of the article as follows: In the next Section we briefly present the Ritus method to derive the Dirac fermion propagator in a general static external magnetic field. In Sect.~\ref{sec3} we present the first-order intertwining framework to generate further inhomogeneous magnetic field profiles from seed (known) solutions to the Ritus eigenfunctions. We  work out the explicit examples of non-trivial magnetic fields derived from the seed uniform and the exponentially decaying magnetic field Ritus eigenfunctions in detail. In Sect.~\ref{sec4} we  derive the electric charge and current densities from the constructed propagator. Finally, we conclude in Sect.~\ref{sec5}. 

\section{Fermion propagator in external magnetic fields}\label{sec2}
We start our discussion of the construction of the fermion propagator in external magnetic fields within the Ritus formalism (see Ref.~\cite{Murguia2010} for a pedagogical presentation of the framework). Such a construction is relevant for monolayer graphene and other 2D materials for which the charge carriers behave as Dirac fermions. Let us consider a magnetic field  pointing perpendicularly to the plane of motion of Dirac fermions, in such a way that, working in a Landau-like gauge, we introduce an electromagnetic potential  $A^{\mu}=(0,0, \mathcal{W}_{0}(x))$~\footnote{In our conventions, Greek indices $\mu, \nu,  \ldots=$0,~1,~2, whereas Latin indices $i, j, \ldots=$1,~2.}, where $\mathcal{W}_{0}(x)$ is a scalar function such that $\mathcal{W}_{0}'(x)=\partial_x \mathcal{W}_{0}(x)$ defines the profile of the field.
In these circumstances, the fermion propagator cannot be diagonalized on the basis of the kinetic momentum eigenfunctions, because the asymptotic states of these fermions in a background magnetic fields do not correspond to plane waves. Motivated by this observation, we notice that  the Green function for Dirac particles, $G(z,z')$, satisfies
\be
((\gamma \cdot \Pi) -m)G(z,z')=\delta^{(3)}(z -z'), \label{Green function}
\ee
with $z^\mu=(t,x,y)$, $\gamma^\mu$ denoting the Dirac matrices (we consider the representation $\gamma^0 = \sigma_3$, $\gamma_1=i\sigma_1$, and $\gamma^2=i\sigma_2$ where $\sigma_i$ are the Pauli matrices), and $\Pi_\mu$ is the canonical momentum. We omit the Lorentz index in the vectors to keep a shorthand notation when necessary. Moreover, although for monolayer graphene the mass gap $m$ vanishes, it becomes a relevant parameter in other systems, and that is why we keep it finite. Eventually, we discuss the limit $m\to 0$. Since $G(z,z')$ commutes with $(\gamma \cdot \Pi)^2$, we expand the propagator on the basis of the eigenfuctions of the later, namely,
the functions $\mathbb{E}_p(z)$ satisfying
\be
(\gamma \cdot \Pi)^2\mathbb{E}_p(z)=p^2\mathbb{E}_p(z), \label{Ritus}
\ee
where the eigenvalue $p^2$ can be any real number corresponding, as we shortly will see, to the magnitude squared of the vector $p^\mu$ (or simply $p$ to avoid cumbersome notation) that labels the functions $\mathbb{E}_p(z)$. We refer to the functions $\mathbb{E}_p(z)$ as the Ritus eigenfunctions~\cite{RITUS1972,Ritus1974,Ritus1978}.
It can be directly verified that these functions $\mathbb{E}_p(z)$ fulfill the closure and completeness relations
\begin{subequations}
\begin{align}\label{orthogonality}
\int \mathrm{d}^3z\, \Bar{\mathbb{E}}_{p'}(z)\mathbb{E}_p(z)&=\mathbb{I}\,\delta(p-p'), \\
\int \mathrm{d}^3p\, \mathbb{E}_p(z')\Bar{\mathbb{E}}_p(z)&=\mathbb{I}\,\delta(z-z'),
\end{align}
\end{subequations}
with $\Bar{\mathbb{E}}_p(z)=\gamma^0\mathbb{E}^{*}_{p}(z)\gamma^0$ and $\mathbb{I}$ is the $2\times 2$ unit matrix.  

In order to construct the Ritus eigenfunctions,  we notice that the operator
\be
(\gamma \cdot \Pi)^2=\gamma^{\mu}\gamma^{\nu}\Pi_{\mu}\Pi_{\nu}=\Pi^2 + \frac{e}{2}\sigma^{\mu\nu}F_{\mu\nu},
\ee
where $F_{\mu\nu}=\partial_\mu A_\nu - \partial_\nu A_\mu$ is the electromagnetic field strength tensor and $\sigma^{\mu\nu}=i[\gamma^\mu, \gamma^\nu]/2$. For a static magnetic field pointing perpendicularly to the plane, the only non-vanishing components of these tensors are
\be
F_{12}=-F_{21}=\mathcal{W}_{0}'(x), \quad \sigma^{12}=\sigma_{3}.
\ee
Then, the eigenvalue Eq.~(\ref{Ritus}) becomes 
\be
(\Pi^{2}+e\sigma_{3}\mathcal{W}_{0}'(x))\mathbb{E}_p(z)=p^2\mathbb{E}_p(z),
\ee
from where we observe that the Ritus eigenfunctions are actually matrices, whose explicit form is
\be
\mathbb{E}_p(z)=\left(\begin{array}{cc}
E_{p,+1}(z) & 0\\
0 & E_{p,-1}(z)
\end{array}\right). \label{ec. eigenvalores}
\ee
Notice that the subscript $p$, which is the shorthand notation of the vector $p^\mu=(p_0,p_2,k)$ is a vector that contains the eigenvalues of the operators $i\partial_t$, $-i\partial_y$, and $\mathcal{H}_{\sigma}$, respectively, and whose norm squared corresponds to the eigenvalue in Eq.~(\ref{Ritus}). That is, the components of the vector $p^\mu$ are the numbers such that
\be
i\partial_t\mathbb{E}_p(z)=p_0\mathbb{E}_p(z), \quad i\partial_y\mathbb{E}_p(z)=-p_2\mathbb{E}_p(z), \quad \mathcal{H}_{\sigma}\mathbb{E}_p(z)=k\mathbb{E}_p(z),
\ee
with $\mathcal{H}_{\sigma}=-(\gamma \cdot \Pi)^2 + \Pi^{2}_{0}$. These eigenvalues allow us to write the scalar functions as
\be
E_{p,\sigma}(z)=e^{-i(p_0t-p_2y)}F_{k,p_2,\sigma}(x), \label{funciones de Ritus}
\ee
where $\sigma=\pm1$ are the eigenvalues of $\sigma_3$ and the functions $F_{k,p_2,\sigma}(x)$ satisfy
\be
[-\partial_{x}^{2} + (p_2 + e\mathcal{W}_{0}(x))^2 - e\sigma \mathcal{W}_{0}'(x) ]F_{k,p_2,\sigma}=kF_{k,p_2,\sigma}, \label{Pauli Hamiltonian}
\ee
which corresponds to a Pauli equation for a particle with mass $m=1/2$ and gyromagnetic factor $g=2$. This equation possesses a supersymmetric structure as we will briefly discuss below. Thus, $F_{k,p_2,\sigma}(x)$ are the solutions of the  equations in~(\ref{Pauli Hamiltonian}) associated to each of the supersymmetric-partner potentials
\be
V_0^{\sigma}(x)=(p_2 + e\mathcal{W}_{0}(x))^2 -e\sigma \mathcal{W}_{0}'(x).
\ee
From now on, we fix the value $\sigma=1$. Then, we have the required ingredients to construct the Ritus eigenfunctions from a first-order supersymmetric formalism.

\section{Supersymmetric framework for the Ritus eigenfunctions}\label{sec3}
Similar to the case of the standard harmonic oscillator, the formalism of first-order supersymmetric quantum mechanics (1-SUSY QM) introduces two first-order differential operators $L_{0}^{\pm}$ explicitly given by
\be
L^{\pm}_{0}=\mp\frac{{\rm d}}{{\rm d}x} + W_{0}(x),
\ee
where $W_{0}(x)$ is known as the superpotential. Here, $L_{0}^{+}$ and $L_{0}^{-}$ are adjoint operators to each other. With them, a pair of Hamiltonians $\mathcal{H}_{+}$ and $\mathcal{H}_{-}$, whose respectively spectra are $k_{n}^{+}$ and $k_{n}^{-}$, can be factorized as
\be
\mathcal{H}_{\pm}=L_{0}^{\pm}L_{0}^{\mp}.
\ee
Here, the so-called intertwining operators $L_{0}^{\pm}$ satisfy the relations
\be 
\mathcal{H}_{\pm}L_{0}^{\pm}=L_{0}^{\pm}\mathcal{H}_{\mp}.
\ee

\begin{figure}[ht]
	\centering
		\includegraphics[width=\textwidth]{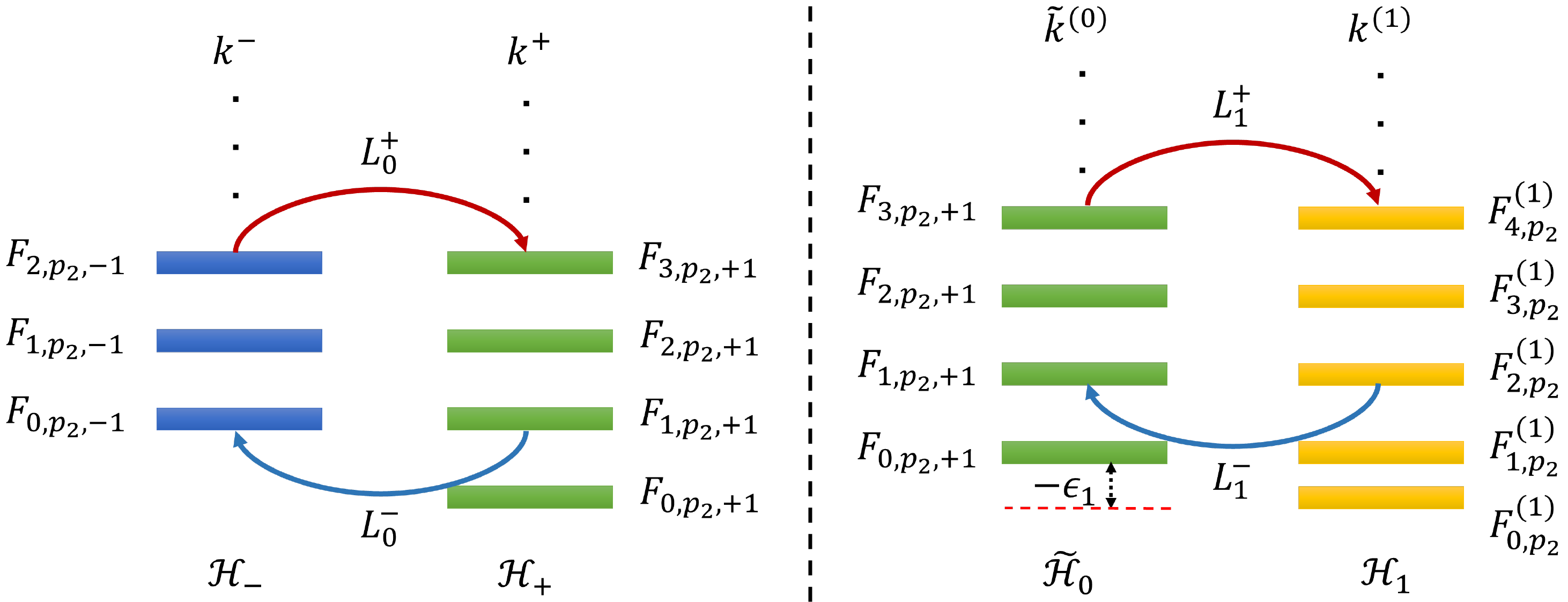}
	\caption{\label{fig:ladders}Representation of eigenfunctions $F_{k,p_{2},\sigma}$ and $F_{k,p_{2}}^{(1)}$, and their corresponding energy levels for the chain of Hamiltonians. The intertwining relationship between the states in Eq.~(\ref{funciones de Ritus}) is shown on the left, while the emergent relationship between the states $F_{k,p_{2},+1}$ and the new functions $F_{k,p_{2}}^{(1)}$ is shown on the right.}
\end{figure}

By a simple inspection, one can recognize that in the construction presented in the  previous section, the functions $F_{k,p_{2},\sigma}$ are related by a supersymmetric transformation. Indeed, the action of the intertwining operators $L_0^{\pm}$ on the solutions of the Hamiltonians~(\ref{Pauli Hamiltonian}) is (see the left panel in Fig.~\ref{fig:ladders})
\be
F_{n,p_2,-1}(x)=\frac{L_0^{-}F_{n+1,p_2,+1}(x)}{\sqrt{k^{+}_{n+1}}},\qquad
F_{n +1,p_2,+1}(x)=\frac{L_0^{+}F_{n,p_2,-1}(x)}{\sqrt{k^{-}_{n}}}, 
\ee
where the ground state, which is annihilated by the operator $L_{0}^{-}$, behaves as
\be
F_{0,p_2,+1}(x)\sim e^{-\int W_{0}(x){\rm d}x}.
\ee
This observation implies that we can write
\be
W_{0}(x)=-\frac{(F_{0,p_2,+1}(x))'}{F_{0,p_2,+1}(x)}.
\ee
Furthermore, the energy levels of $\mathcal{H}_{\pm}$ turn out to be 
\be
k_n^{-}=k_{n+1}^{+}\;, \quad k_{0}^{+}=0.
\ee

These expressions indicate that the eigenfunctions and eigenvalues of the problem can be found through the operators $L_{0}^{\pm}$, which simplifies the calculations since this involves just first-order derivatives. Also, the magnetic field profile $B_0(x)$ can be related to the electromagnetic potential $A_\mu(x)$, the superpotential $W_{0}(x)$, and the ground state of $H^{+}$ as follows:
\be
B_{0}(x)=A_2'(x)=\frac{1}{e}W_{0}'(x)=-\frac{1}{e}\frac{{\rm d}^2}{{\rm d}x^2}\{\ln[F_{0,p_2,+1}(x)]\},
\ee
which implies that is valid to make $\mathcal{W}_{0}(x)\equiv W_{0}(x)$.

\subsection{Generalized first order intertwining}
In this section we introduce the first order supersymmetric formalism to generate inhomogeneous magnetic fields from intertwining operators. We follow closely the discussion of Ref.~\cite{Celeita2020}. Taking as starting Hamiltonian the one with $\sigma=1$ in~(\ref{Pauli Hamiltonian}), the first step of the method consists in displacing the energy of the Hamiltonian $\mathcal{H}_{+}$ as follows:
\be
    \tilde{\mathcal{H}}_{0}\equiv\mathcal{H}_{+}-\epsilon_{1}=-\frac{{\rm d}^2}{{\rm d}x^2}+V_{+}(x)-\epsilon_{1},
\ee
so that $\tilde{V}_{0}(x)=V_{+}(x)-\epsilon_{1}$, where $\epsilon_{1}\leq  k_{0}^{+}=0$. Here $\tilde{\mathcal{H}}_{0}$ is the Hamiltonian upon which the 1-SUSY QM formalism will be applied.

The second step is to build a new Hamiltonian $\mathcal{H}_{1}$ departing from $\tilde{\mathcal{H}}_{0}$ through the intertwining relation (see the right panel in Fig.~\ref{fig:ladders}):
\be
    \mathcal{H}_{1}L_{1}^{+}=L_{1}^{+}\tilde{\mathcal{H}}_{0},
\ee
where $\mathcal{H}_{1}$ and $L_{1}^{\pm}$ are given by
\be
    \mathcal{H}_{1}=-\frac{{\rm d}^2}{{\rm d}x^2}+V_{1}(x,\epsilon_{1}), \quad L_{1}^{\pm}=\mp\frac{{\rm d}}{{\rm d}x}+W_{1}(x,\epsilon_{1}),
\ee
respectively, which implies $\tilde{\mathcal{H}}_{0}=L_{1}^{-}L_{1}^{+}$ and $\mathcal{H}_{1}=L_{1}^{+}L_{1}^{-}$. This leads to the following  relations for $W_1$ and $V_1$ derived from $V_0$,
\begin{subequations}
\begin{align}
&W_{1}^{2}(x,\epsilon_{1})+W'_{1}(x,\epsilon_{1})=\tilde{V}_{0}(x), \\
&V_{1}(x,\epsilon_{1})=\tilde{V}_{0}(x)-2W'_{1}(x,\epsilon_{1}).
\end{align}
\end{subequations}

Let us suppose now that we can write $W_{1}(x,\epsilon_{1})=u'_{1}/u_{1}$.  The above relations lead us to the following expression for $u_1$:
\be
-u''_{1}+\tilde{V}_{0}(x)u_{1}=0. \label{24}
\ee
The corresponding magnetic field giving place to $V_{1}(x,\epsilon_{1})$ is obtained from
\be
B_{1}(x,\epsilon_{1})=\frac{ 1}{e}\frac{d\,W_{1}(x,\epsilon_{1})}{dx}=-B_{0}(x)-\frac{1}{e}\frac{{\rm d}^2}{{\rm d}x^2}\left\{\ln\left[\frac{F_{0,p_2,+1}(x)}{u_{1}}\right]\right\}. \label{25}
\ee

The third step of the method is to identify the eigenfunctions and eigenvalues of the new system. The energy levels for $\tilde{\mathcal{H}}_{0}$ and $\mathcal{H}_{1}$ are those of $\mathcal{H}_{+}$, displaced by the quantity $-\epsilon_{1}$, plus the ground state of $\mathcal{H}_{1}$ at zero energy:
\begin{subequations}
\begin{align}
\tilde{k}_{n}^{(0)}&=k_{n}^{+}-\epsilon_{1}, \\
k_{0}^{(1)}&=0, \quad k_{n+1}^{(1)}=\tilde{k}_{n}^{(0)}, \quad n=0,1,\dots,
\end{align}
\end{subequations}
with $\epsilon_{1}\leq k_{0}^{+}=0$. The unknown eigenfunctions associated with these energies are given by:
\be
F_{0,p_{2}}^{(1)}(x)\sim\frac{1}{u_{1}}, \qquad
F^{(1)}_{n+1,p_{2}}(x)=\frac{1}{\sqrt{\tilde{k}_{n}^{(0)}}}L_{1}^{+}F_{n,p_{2},+1}(x).
\ee
where the eigenfunctions $F_{n,p_{2},+1}(x)$ of $\mathcal{H}_{+}$, and consequently those of $\tilde{\mathcal{H}}_{0}$, are assumed to be known. In addition, the ground state of $\mathcal{H}_{1}$ fulfills the condition $L_{1}^{-}F_{0,p_{2}}^{(1)}(x)=0$.

It is worth noting that, according to the 1-SUSY QM formalism, since $\epsilon_{1}\leq k_{0}^{+}$ and depending on the choice of the function $u_{1}$, three different cases can arise for the spectrum of the Hamiltonian $\mathcal{H}_{1}$: that it does not include the ground state of $\tilde{\mathcal{H}}_{0}$, or that it has an extra energy level, or that it is isospectral to $\tilde{\mathcal{H}}_{0}$. Below we discuss in detail two examples of magnetic field profiles, namely, the homogeneous field and the exponentially decaying magnetic field, only for one such case.

\subsubsection{Uniform magnetic field}
First, let us consider a uniform magnetic field, for which the vector potential is
\be
\mathbf{A}(x)=B_{0}x\,\hat{\mathbf{y}} \quad \Longrightarrow \quad \mathbf{B}_{0}(x)=B_{0}\,\hat{\mathbf{z}},
\ee
and the corresponding superpotential reads as
\be
W_{0}(x)=\frac{\omega}{2}x+p_{2}, \quad \omega=2eB_{0}. \label{Wpotential}
\ee
From this function, we obtain the superpartner potential which give explicitly that
\be
\mathcal{H}_{+}=-\frac{{\rm d}^2}{{\rm d}x^2}+V_{+}(x)=-\frac{{\rm d}^2}{{\rm d}x^2}+\frac{\omega^2}{4}\left(x+\frac{2p_{2}}{\omega}\right)^{2}-\frac{\omega}{2}, \label{HOpotential}
\ee
and its eigenenergies that correspond to those of a shifted quantum harmonic oscillator 
\be
k_{n}^{+}=\omega\,n, \quad n=0, 1, 2, \dots,   \label{spectrumconst}
\ee
while the corresponding eigenfunctions can be expressed as
\be
F_{n,p_{2},+1}(x)=N_{n}e^{-\frac{\omega}{4}\left(x+\frac{2p_{2}}{\omega}\right)^{2}}H_{n}\left[\sqrt{\frac{\omega}{2}}\left(x+\frac{2p_{2}}{\omega}\right)\right],
\ee
 with $N_{n}=\sqrt{\frac{1}{2^{n}n!}\left(\frac{\omega}{2\pi}\right)^{1/2}}$ being the normalization constant  and $H_{n}(x)$ are the Hermite polynomials.  By defining the dimensionless quantity
\be
\eta(x)\equiv\sqrt{\frac{\omega}{2}}\left(x + \frac{2p_2}{\omega}\right)\;,
\ee
we simplify the eigenfunctions as
\be
F_{n,p_{2},+1}(\eta)=N_{n}e^{-\eta^{2}/2}H_{n}\left(\eta\right).
\ee
This expression corresponds to our seed solution.
Next, we want to construct a  non-trivial magnetic field profile starting from the uniform case by applying the 1-SUSY QM formalism.
As stated earlier, the first step is to shift the energy of $\mathcal{H}_{+}$ as follows:
\be
\tilde{\mathcal{H}}_{0}=\mathcal{H}_{+}-\epsilon_{1}=-\frac{\omega}{2}\frac{{\rm d}^2}{{\rm d}\eta^2}+\frac{\omega}{2}\,\eta^2-\frac{\omega}{2}-\epsilon_{1},
\ee
with $\epsilon_{1}\leq k_{0}^{+}=0$. Thus, the potential $\tilde{V}_{0}$ reads
\be
\tilde{V}_{0}(\eta)=\frac{\omega}{2}\,\eta^2-\frac{\omega}{2}-\epsilon_{1}.
\ee

\begin{figure}[ht]
	\centering
	\begin{minipage}[b]{0.48\textwidth}
		\includegraphics[width=\textwidth]{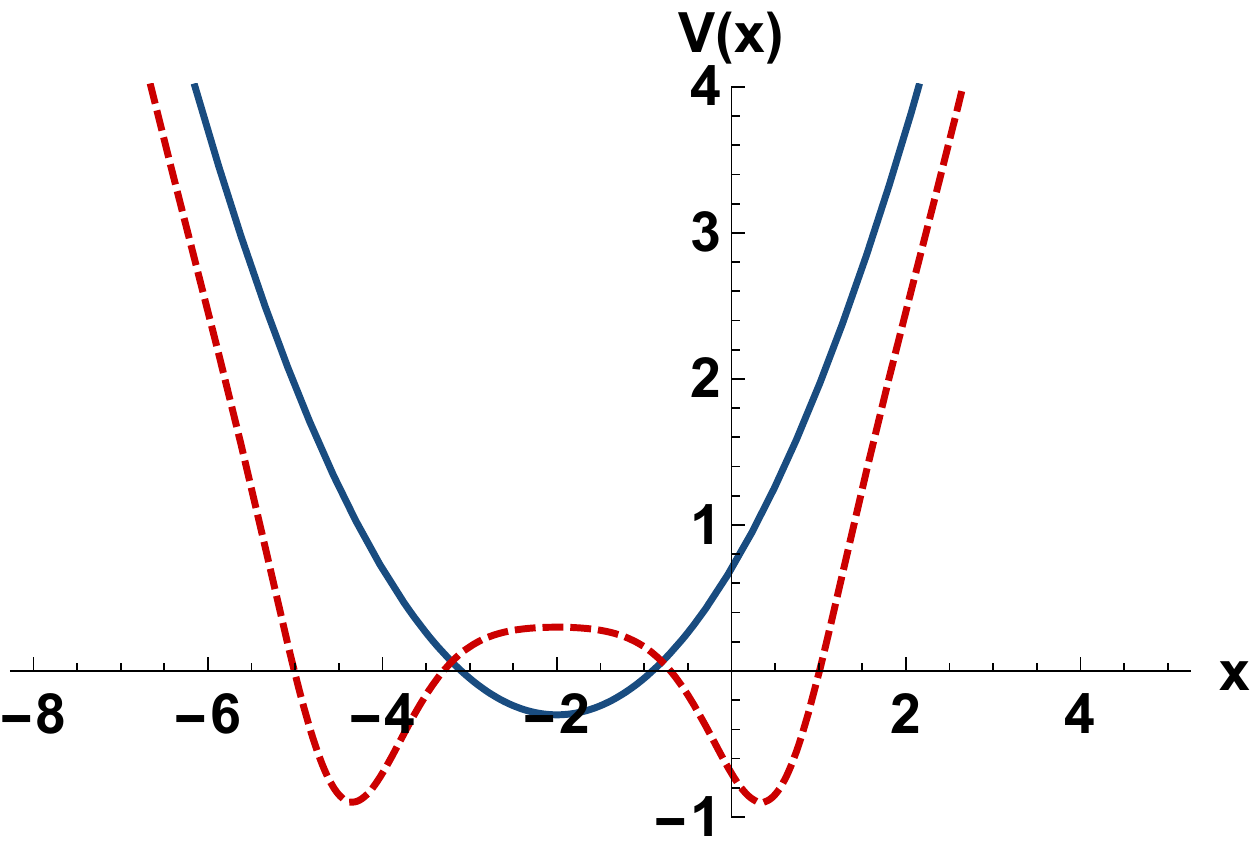}\\
		\centering{\footnotesize (a)}
		\label{fig:fig1a}
	\end{minipage}
	\hspace{0.25cm}
	\begin{minipage}[b]{0.48\textwidth}
		\includegraphics[width=\textwidth]{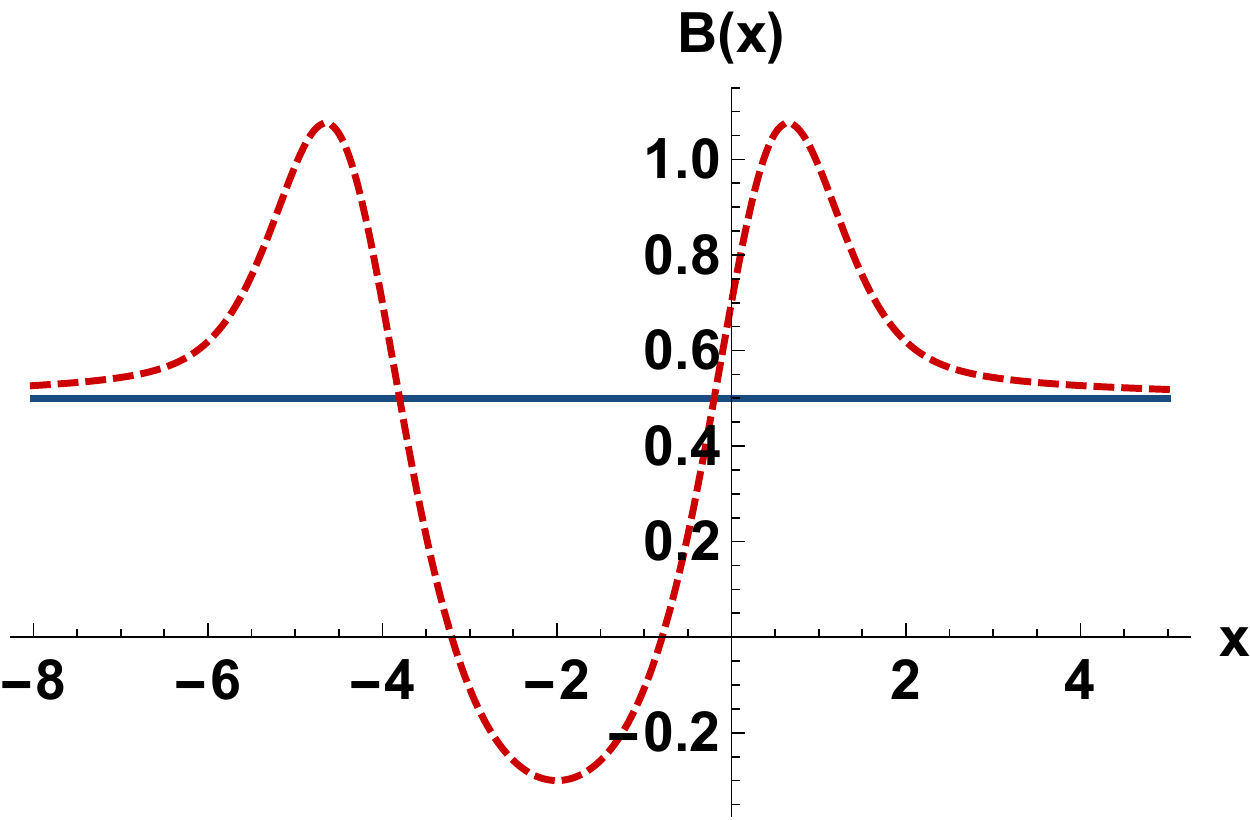}\\
		\centering{\footnotesize (b)}
		\label{fig:fig1b}
	\end{minipage}
	\caption{\label{fig:fig1}(a) Generated potential $V_1(x,\epsilon_1)$ (red, \dashed) and the initial one $\tilde{V}_0(x)$ (dark blue, \full). (b) Generated magnetic field $B_1(x, \epsilon_1)$ (red, \dashed) and the constant initial one $B_0$ (dark blue, \full). In both cases $B_{0}=\frac12$, $p_{2}=1$, $\epsilon_{1}=-\frac{\omega}{5}$ and $\omega = 1$.}
\end{figure}

From here, we can readily obtain $W_1(x,\epsilon_1)$ and correspondingly, $V_1(x,\epsilon_1)$. Then, from the replacement $W_{1}(x,\epsilon_{1})=u'_{1}/u_{1}$ in Eq.~(\ref{24}), we easily infer that
\begin{equation}
u_{1}=e^{-\eta^{2}/2}\Bigg(\,_{1}F_{1}\left[a,\frac{1}{2},\eta^{2}\right]+2\nu_{1}\frac{\Gamma(a+1/2)}{\Gamma(a)}\eta\,_{1}F_{1}\left[a+\frac{1}{2},\frac{3}{2},\eta^{2}\right]\Bigg),
\end{equation}
with $a=-\epsilon_{1}/(2\omega)$, $\nu_{1}\in(-1,1)$. For definitiveness and comparison with the findings of Ref.~\cite{Celeita2020}, by choosing the parameters $\epsilon_{1}=-k_{1}^{+}/5=-\omega/5$ and $\nu_{1}=0$, we have $a=1/10$ and
\begin{subequations}
\begin{align}
W_{1}(\eta,\epsilon_{1})&=\sqrt{\frac{\omega}{2}}\eta\left(-1 +\frac{2}{5}\frac{\,_{1}F_{1}\left[\frac{11}{10},\frac{3}{2},\eta^2\right]}{\,_{1}F_{1}\left[\frac{1}{10},\frac{1}{2},\eta^2\right]}\right), \\
V_{1}(\eta,\epsilon_{1})&=\tilde{V}_{0}(\eta)-\sqrt{2\omega}\frac{{\rm d}}{{\rm d}\eta}\left[\sqrt{\frac{\omega}{2}}\eta\left(-1+\frac{2}{5}\frac{\,_{1}F_{1}\left[\frac{11}{10},\frac{3}{2},\eta^2\right]}{\,_{1}F_{1}\left[\frac{1}{10},\frac{1}{2},\eta^2\right]}\right)\right], \\
B_{1}(\eta,\epsilon_{1})&=-B_{0}+\frac{2B_{0}}{5}\frac{{\rm d}}{{\rm d}\eta}\left[\sqrt{\frac{2}{\omega}}\eta\frac{\,_{1}F_{1}\left[\frac{11}{10},\frac{3}{2},\eta^2\right]}{\,_{1}F_{1}\left[\frac{1}{10},\frac{1}{2},\eta^2\right]}\right].\label{eq:nontrB1}
\end{align}
\end{subequations}
A plot of the generated potential $V_{1}(x,\epsilon_{1})$ and the magnetic field profile $B_{1}(x,\epsilon_{1})$ in this case is shown in Fig.~\ref{fig:fig1}. Then, the eigenenergies of the system are explicitly
\be
k_{0}^{(1)}=0, \quad k_{n+1}^{(1)}=\omega\left(n+\frac15\right), \quad n=0,1,2,\dots,
\ee
while the corresponding Ritus eigenfunctions, taking into account (\ref{funciones de Ritus}), are given by:
\begin{subequations}
\begin{align}
E_{0,p}^{(1)}(\eta,y,t)&\sim \exp\left(-i(p_{0}t-p_{2}y)\right)\frac{e^{\eta^{2}/2}}{\,_{1}F_{1}\left[\frac{1}{10},\frac{1}{2},\eta^2\right]}, \label{eigenvalor campo constante} \\
E_{n+1,p}^{(1)}(x,y,t)&=\exp\left(-i(p_{0}t-p_{2}y)\right)F_{n+1,p_{2}}^{(1)}(x) \nn \\
&=\exp\left(-i(p_{0}t-p_{2}y)\right)\frac{1}{\sqrt{\omega(n+1/5)}}L_{1}^{+}F_{n,p_{2},+1} \nn \\
&=\frac{\exp\left(-i(p_{0}t-p_{2}y)\right)}{\sqrt{2(n+1/5)}}\left(\frac{2\,\eta}{5}\frac{\,_{1}F_{1}\left[\frac{11}{10},\frac{3}{2},\eta^2\right]}{\,_{1}F_{1}\left[\frac{1}{10},\frac{1}{2},\eta^2\right]}F_{n,p_{2},+1}-\sqrt{2n}F_{n-1,p_{2},+1}\right), \label{eigenvalor campo constante 1}
\end{align}
\end{subequations}
for $n=0,1,2,\dots$. The joint choice of $\epsilon_{1}$ and the function $u_{1}$ allows that the energy spectrum of $\mathcal{H}_{1}$ has an extra level in comparison with that of $\tilde{\mathcal{H}}_{0}$.

Inserting these expressions into Eq.~(\ref{ec. eigenvalores}), we obtain the Ritus eigenfunctions for a seed constant magnetic field for the graphene to first-order intertwining which gives raise to the highly non-trivial magnetic field profile in Eq.~\eqref{eq:nontrB1}.

\subsubsection{Exponentially decaying magnetic field}
Let us now consider the vector potential
\be
\mathbf{A}(x)=-\frac{B_{0}}{\alpha}(\exp\left(-\alpha x\right)-1)\,\hat{\mathbf{y}}, \quad B_{0}>1, \quad \alpha\geq0,
\ee
where we refer to $\alpha$  as the inhomogeneity term.

Thus, we have
\be
\mathbf{B}_{0}(x)=B_{0}\exp\left(-\alpha x\right)\,\hat{\mathbf{z}} \quad \Longrightarrow \quad W_{0}(x)=p_{2}-D(\exp\left(-\alpha x\right)-1), \quad D=\frac{e\,B_{0}}{\alpha}, \label{superpotentialexp}
\ee
which leads to the Morse potentials:
\be
V_{0}^{\sigma}(x)=q_{2}^{2}+D^{2}\exp\left(-2\alpha x\right)-2D\left(q_{2}+\sigma\frac{\alpha}{2}\right)\exp\left(-\alpha x\right), \label{Morsepotential}
\ee
where $q_{2}=p_{2}+D$. Note that our results coincide with those in Refs.~\cite{Celeita2020,knn09} making the replacement $q_{2}\rightarrow k_{2}$.

By defining the quantity 
\begin{equation}
    \rho(x)\equiv\frac{2D}{\alpha}\exp\left(-\alpha x\right),
\end{equation}
the eigenfunctions $F_{n,p_{2},+1}(\rho)$ of $\mathcal{H}_{+}$ are given by~\cite{Celeita2020,knn09,Ghosh_2008}
\be
F_{n,p_{2},+1}(\rho)=N_{n}\exp\left(-\rho/2\right)\rho^{(q_{2}/\alpha-n)}L_{n}^{2(q_{2}/\alpha-n)}(\rho), \quad n=0,1,2,\dots\leq q_{2}/\alpha, \label{eigenfunction exponential}
\ee
where $N_{n}$ is the corresponding normalization constant and $L_{n}^{\alpha}(x)$ are the Laguerre polynomials, and its eigenenergies turn out to be
\be 
k_{n}^{+}=\alpha\,n\left(2 q_{2} - \alpha\,n\right), \quad n=0,1,\dots. \label{spectrumexp}
\ee

We chose $V_{0}^{+}(\rho)$ and displace it by $-\epsilon_{1}$ to produce $\tilde{V}_{0}(\rho)$, namely,
\be
\tilde{V}_{0}(\rho)=q_{2}^{2}+\frac{\alpha^{2}}{4}\rho^{2}-\alpha\rho\left(q_{2}+\frac{\alpha}{2}\right)-\epsilon_{1}.
\ee
Again, the new potential $V_{1}(\rho,\epsilon_{1})$ depends on $W_{1}(\rho,\epsilon_{1})$, which is a solution of the Riccati equation:
\begin{subequations}
\begin{align}
&W_{1}^{2}(\rho,\epsilon_{1})+W'_{1}(\rho,\epsilon_{1})=\tilde{V}_{0}(\rho), \\
&V_{1}(\rho,\epsilon_{1})=\tilde{V}_{0}(\rho)-2W'_{1}(\rho,\epsilon_{1}). \label{48b}
\end{align}
\end{subequations}

\begin{figure}[ht]
	\centering
	\begin{minipage}[b]{0.48\textwidth}
		\includegraphics[width=\textwidth]{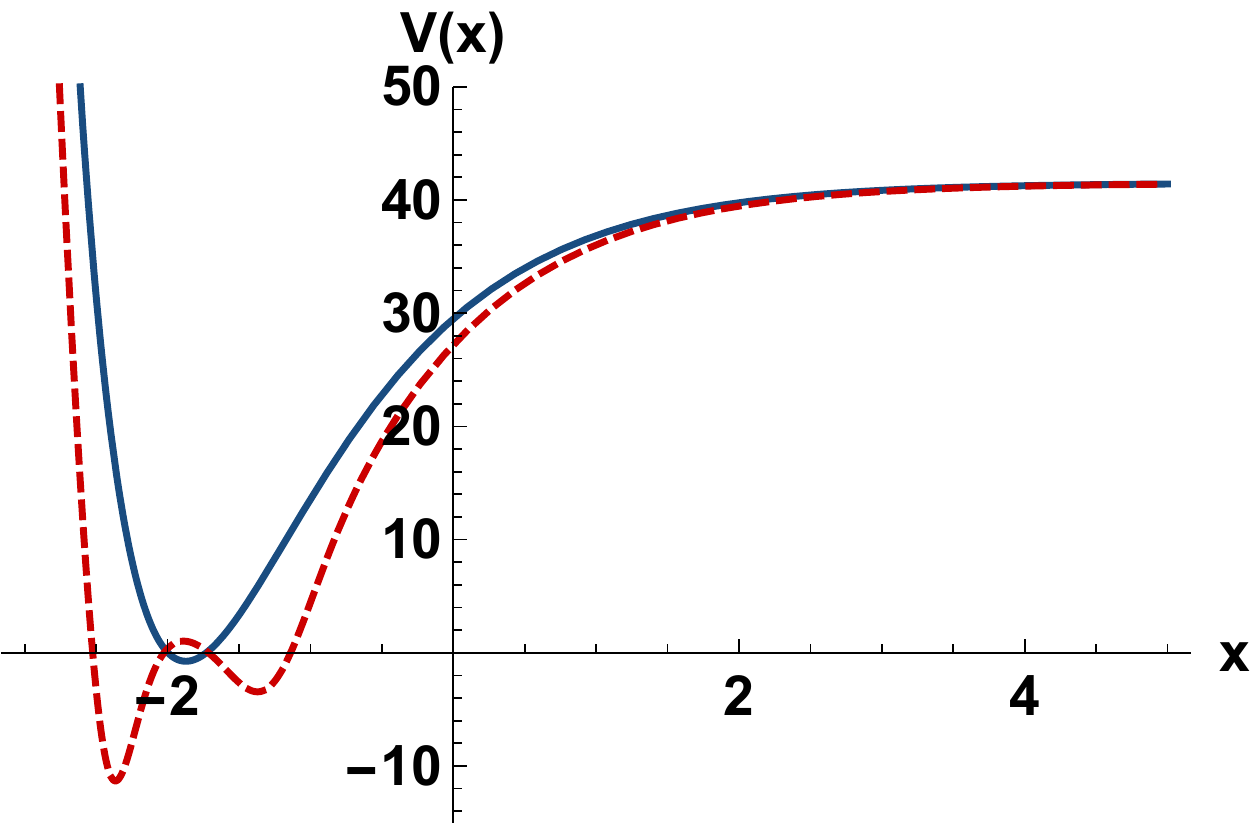}\\
		\centering{\footnotesize (a)}
		\label{fig:fig2a}
	\end{minipage}
	\hspace{0.25cm}
	\begin{minipage}[b]{0.48\textwidth}
		\includegraphics[width=\textwidth]{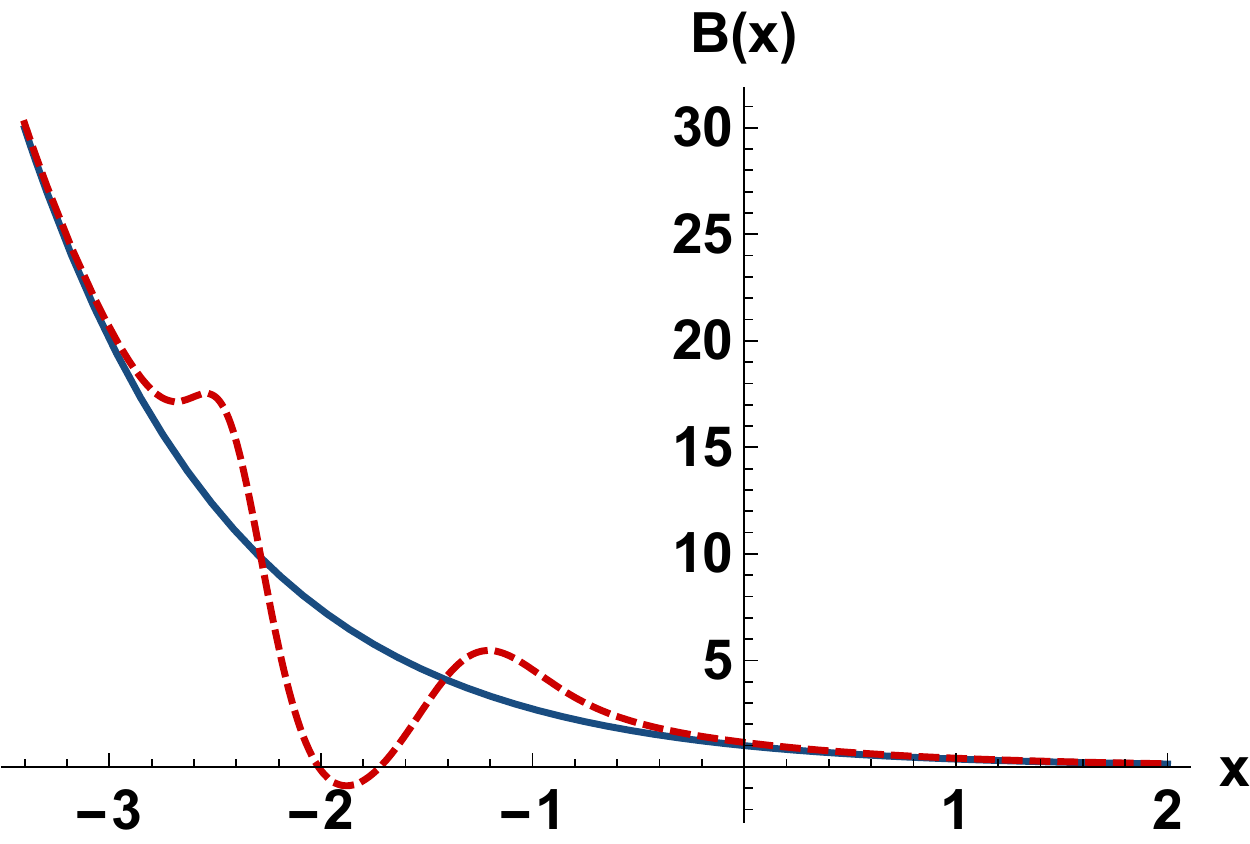}\\
		\centering{\footnotesize (b)}
		\label{fig:fig2b}
	\end{minipage}
	\caption{\label{fig:fig2}(a) Generated potential $V_1(x,\epsilon_1)$ (red, \dashed) and the initial one $\tilde{V}_0(x)$ (dark blue, \full). (b) Generated magnetic field $B_1(x, \epsilon_1)$ (red, \dashed) and the initially decaying magnetic field (dark blue, \full). In both cases $B_{0}=1$, $\nu_{1}=-\frac32$, $p_{2}=5\alpha$, $\epsilon_{1}=-\frac{11\alpha^{2}}{2}$ and $\alpha=1$.}
\end{figure}

The new superpotential is written as $W_{1}(\rho,\epsilon_{1})=u'_{1}/u_{1}$, with $u_{1}$ being the general solution of the Schr\"odinger equation
\begin{subequations}
\begin{align}
&-u''_{1}+\tilde{V}_{0}(\rho)u_{1}=0,\\
u_{1}&=\exp\left(-\rho/2\right)\left(\frac{\alpha\rho}{2D}\right)^{\sqrt{q_{2}^{2}-\epsilon_{1}}/\alpha}\left(\,_{1}F_{1}\left[a,b,\rho\right]+\left(\frac{2q_{2}}{\alpha}\right)\left(1+\frac{1}{\nu_{1}}\right)U\left[a,b,\rho\right]\right),
\end{align}
\end{subequations}
where $\nu_{1}$ obeys the restriction $\nu_{1}\in\mathbb{R}-\{[-1,0]\}$ and the parameters $a$ and $b$ are defined as: 
\be
a=-\frac{q_{2}}{\alpha}+\frac{\sqrt{q_{2}^{2}-\epsilon_{1}}}{\alpha}, \quad b=1+\frac{2\sqrt{q_{2}^{2}-\epsilon_{1}}}{\alpha}.
\ee
Therefore, the superpotential turns out to be:
\be
W_{1}(\rho,\epsilon_{1})=\frac{\alpha\,\rho}{2}-\sqrt{q_{2}^{2}-\epsilon_{1}}+\mathcal{F}(\rho),
\ee
where the function $\mathcal{F}(\rho)$ reads
\begin{align}
\mathcal{F}(\rho)&=-\left(\frac{\alpha\,a}{b}\rho\right)\frac{\,_{1}F_{1}\left[1+a,1+b,\rho\right]-\frac{2q_{2}b}{\alpha}\left(1+\frac{1}{\nu_{1}}\right)U\left[1+a,1+b,\rho\right]}{\,_{1}F_{1}\left[a,b,\rho\right]+\frac{2q_{2}}{\alpha}\left(1+\frac{1}{\nu_{1}}\right)U\left[a,b,\rho\right]}.
\end{align}

Thus, the new potential and associated magnetic field are given by (see Eqs.~(\ref{25}) and~(\ref{48b})):
\begin{subequations}
\begin{align}
V_{1}(\rho,\epsilon_{1})&=\tilde{V}_{0}(\rho)+2\alpha\rho\frac{{\rm d}}{{\rm d}\rho}\left[\mathcal{F}(\rho)+\frac{\alpha\rho}{2}\right], \label{SUSYpotexp} \\
B_{1}(\rho,\epsilon_{1})&=-\frac{\alpha^{2}\rho}{2e}-\frac{\alpha\rho}{e}\frac{{\rm d}}{{\rm d}\rho}\mathcal{F}(\rho). \label{SUSYmagnexp}
\end{align}
\end{subequations}
A plot of the generated potential $V_{1}(x,\epsilon_{1})$ and the magnetic field profile $B_{1}(x,\epsilon_{1})$ in this case is shown in Fig.~\ref{fig:fig2}.

In order to compare our results, we set the factorization energy as $\epsilon_{1}=-k_{1}^{+}/2=-\alpha(2q_{2}-\alpha)/2$, so the eigenenergies for the problem are given by:
\be
k_{0}^{(1)}=0, \quad k_{n+1}^{(1)}=\tilde{k}_{n}^{(0)}=\alpha\,n(2q_{2}-\alpha\,n)+\frac{\alpha}{2}(2q_{2}-\alpha), \quad n=0,1,\dots.
\ee
The eigenfunctions corresponding to $\mathcal{H}_{1}$ take the form
\begin{subequations}
\begin{align}
F_{0,p_{2}}^{(1)}(\rho)&\sim\frac{\exp\left(\rho/2\right)\left(\frac{2D}{\alpha\rho}\right)^{\sqrt{q_{2}^{2}-\epsilon_{1}}/\alpha}}{\,_{1}F_{1}\left[a,b,\rho\right]+\left(\frac{2q_{2}}{\alpha}\right)\left(1+\frac{1}{\nu_{1}}\right)U\left[a,b,\rho\right]}, \\
F_{n+1,p_{2}}^{(1)}(\rho)&=\frac{1}{\sqrt{\alpha\left[n(2q_{2}-\alpha\,n)+\frac{1}{2}(2q_{2}-\alpha)\right]}}L_{1}^{+}(\rho,\epsilon_{1})F_{n,p_{2},+1}(\rho) \nonumber \\
&=\frac{1}{\sqrt{\alpha\left[n(2q_{2}-\alpha\,n)+\frac{1}{2}(2q_{2}-\alpha)\right]}} \nonumber \\
&\quad\times\left(\left(q_{2}-\sqrt{q_{2}^{2}-\epsilon_{1}}+\mathcal{F}(\rho)\right)-A^{-}\right)F_{n,p_{2},+1}(\rho),
\end{align}
\end{subequations}
for $n=0,1,2,\dots$, where
\begin{equation}
    A^{-}=-\alpha\rho\frac{{\rm d}}{{\rm d}\rho}+\left(q_{2}-\frac{\alpha\rho}{2}\right).
\end{equation}
Therefore, the corresponding Ritus eigenfunctions, taking into account (\ref{funciones de Ritus}), are given by:
\begin{subequations}
\begin{align}
E_{0,p}^{(1)}(\rho,y,t)&=\exp\left(-i(p_{0}t-p_{2}y)\right)F_{0,p_2}^{(1)}(\rho)  \nn \\
&\sim \exp\left(-i(p_{0}t-p_{2}y)\right)\frac{\exp\left(\rho/2\right)\left(\frac{2D}{\alpha\rho}\right)^{\sqrt{q_{2}^{2}-\epsilon_{1}}/\alpha}}{\,_{1}F_{1}\left[a,b,\rho\right]+\left(\frac{2q_{2}}{\alpha}\right)\left(1+\frac{1}{\nu_{1}}\right)U\left[a,b,\rho\right]}, \label{eigenvalor exponencial}\\
E_{n+1,p}^{(1)}(x,y,t)&=\exp\left(-i(p_{0}t-p_{2}y)\right)F_{n+1,p_{2}}^{(1)}(x) \nn \\
&=\exp\left(-i(p_{0}t-p_{2}y)\right)\frac{\left(\left(q_{2}-\sqrt{q_{2}^{2}-\epsilon_{1}}+\mathcal{F}(\rho)\right)-A^{-}\right)F_{n,p_{2},+1}(\rho)}{\sqrt{\alpha\left[n(2q_{2}-\alpha n)+\frac{1}{2}(2q_{2}-\alpha)\right]}}, \label{eigenvalor exponencial 1}
\end{align}
\end{subequations}
for $n=0,1,2,\dots$. Once again, the joint selection of $\epsilon_{1}$ and $u_{1}$ allows that $\mathcal{H}_{1}$ has an extra energy level than $\tilde{\mathcal{H}}_{0}$.

Inserting these expressions into Eq. (\ref{ec. eigenvalores}), we obtain the Ritus eigenfunctions for a seed exponentially decaying static magnetic field for the graphene to first-order intertwining in the magnetic field \eqref{SUSYmagnexp}.

\section{Charge and current density}\label{sec4}
Physically, Ritus eigenfunctions $\mathbb{E}_p(z)$ correspond to the asymptotic states of electrons in graphene with momentum $\bar{p}$ in the external field. Therefore, we can use these functions to diagonalize the fermion propagator  $S(z,z')$ in momentum space in the same way plane waves are used to define the Fourier transform, 
\be
S(z,z')=\int {\rm d}^3p\,\mathbb{E}_p(z)\ S_F(p)\ \Bar{\mathbb{E}}_{p'}(z')\;.
\ee
Inserting this Green's functions in Eq. (\ref{Green function}), using the property \cite{Cooper1995}
\be
(\gamma \cdot \Pi)\mathbb{E}_p(z)=\mathbb{E}_p(z)(\gamma \cdot \bar{p})\;,
\ee
where $\bar{p}$ is the shorthand notation to define the three-momentum vector  $\bar{p}^\mu=(p_0, 0, \sqrt{k})$ that satisfies $\bar{p}^{2}=p^2=p_0^{2} -k$ \cite{Murguia2010} and the properties (\ref{orthogonality}), the propagator in momentum space takes the form 
\be
S_{F}(p)=\frac{1}{\gamma\cdot\Bar{p}-m},
\ee
similar to the free-particle propagator, but the momentum $\bar{p}$, which carries the quantum numbers induced on the dynamics of Dirac fermions by in the presence of the external field. In the configuration space, we write the propagator as
\begin{align}
S(z,z')&=\int {\rm d}^{3}p\ \mathbb{E}_{p}(z)\left[\frac{1}{\gamma\cdot\Bar{p}-m}\right]\Bar{\mathbb{E}}_{p'}(z')\nn\\
&=\int {\rm d}^{3}p\ \mathbb{E}_{p}(z)\left[\frac{\gamma\cdot\Bar{p}+m}{p^{2}-m^{2}}\right]\Bar{\mathbb{E}}_{p'}(z'). 
\end{align}
From this expression we can find the value of the electric charge and induced vacuum current densities. First, we notice that we can write 
\be
\mathbb{E}_{p}=E_{p,+1}P_{+}+E_{p,-1}P_{-}, \qquad  \Bar{\mathbb{E}}_{p}=E_{p,+1}^{\ast}P_{+}+E_{p,-1}^{\ast}P_{-},
\ee
where we use projection operators $P_{\pm}=\frac12(\mathbb{I}\pm\gamma^{0})=\frac12(\mathbb{I}\pm i\gamma^{1}\gamma^{2})$, or more explicitly:
\be
P_{+}=\left(\begin{array}{cc}
    1 & 0 \\
    0 & 0
\end{array}\right), \quad P_{-}=\left(\begin{array}{cc}
    0 & 0 \\
    0 & 1
\end{array}\right),
\ee
which satisfy $P_{\pm}P_{\pm}=P_{\pm}$ and $P_{\pm}P_{\mp}=0$. Hence, from the definition
\be
j^\mu= {\rm Tr}\{ \gamma^\mu S(z,z')\},
\ee
we have that
\begin{align}
{\rm Tr}\{\gamma^{0}S(z,z')\}&={\rm Tr}\left\{\gamma^{0}\int {\rm d}^{3}p\left[\frac{\gamma^{\mu}p_{\mu}+m}{p^{2}-m^{2}}\right](\vert E_{p,+1}\vert^{2}P_{+}+\vert E_{p,-1}\vert^{2}P_{-})\right\} \nn \\
&=\int \frac{{\rm d}^{3}p}{p^{2}-m^{2}}\left({p_{\mu}\rm Tr}\left\{\frac{\gamma^{0}\gamma^{\mu}}{2}(\vert E_{p,+1}\vert^{2}+\vert E_{p,-1}\vert^{2})+\frac{\gamma^{0}\gamma^{\mu}\gamma^{0}}{2}(\vert E_{p,+1}\vert^{2}+\vert E_{p,-1}\vert^{2})\right\}\right.\nn \\
&\quad\left.+m\,{\rm Tr}\left\{\frac{\gamma^{0}}{2}(\vert E_{p,+1}\vert^{2}+\vert E_{p,-1}\vert^{2})+\frac{\gamma^{0}\gamma^{0}}{2}(\vert E_{p,+1}\vert^{2}+\vert E_{p,-1}\vert^{2})\right\}\right).
\end{align}
Upon taking traces, we obtain
\be
{\rm Tr}\{\gamma^{0}S(z,z')\}=\int\frac{{\rm d}^{3}p}{p^{2}-m^2}\left(p_{0}(\vert E_{p,+1}\vert^{2}+\vert E_{p,-1}\vert^{2})+ m(\vert E_{p,+1}\vert^{2}+\vert E_{p,-1}\vert^{2})\right).
\ee
Therefore:
\begin{align}
j^{0}&=-ie\,{\rm Tr}\{\gamma^{0}S(z,z')\} \nn \\ &=-ie\int\frac{{\rm d}^{3}p}{p^{2}-m^2}\Bigg(p_{0}(\vert E_{p,+1}\vert^{2}+\vert E_{p,-1}\vert^{2})+ m(\vert E_{p,+1}\vert^{2}+\vert E_{p,-1}\vert^{2})\Bigg).
\end{align}
Since the first integral is odd respect to $p_{0}$, we have that
\be
j^{0}=-ie\int\frac{{\rm d}^{3}p}{p^{2}-m^2}m(\vert E_{p,+1}\vert^{2}+\vert E_{p,-1}\vert^{2})\;. \label{ec. carga}
\ee

On the other hand,
\be
j^{\ell}=-ie\,{\rm Tr}\{\gamma^{\ell}S(z,z')\}, \quad \ell=1,2,
\ee
where
\begin{align}
{\rm Tr}\{\gamma^{\ell}S(z,z')\}&=\int\frac{{\rm d}^{3}p}{p^{2}-m^{2}}\bigg[\Bar{p}_{\mu}{\rm Tr}\{\gamma^{\ell}(E_{p,+1}P_{+}+E_{p,-1}P_{-})\gamma^{\mu}(E_{p,+1}^{\ast}P_{+}+E_{p,-1}^{\ast}P_{-})\} \nn \\
&\quad+m\,{\rm Tr}\{\gamma^{\ell}(E_{p,+1}P_{+}+E_{p,-1}P_{-})(E_{p,+1}^{\ast}P_{+}+E_{p,-1}^{\ast}P_{-})\}\bigg] \nn \\
&=\int\frac{{\rm d}^{3}p}{p^{2}-m^{2}}\Bigg[\Bar{p}_{\mu}{\rm Tr}\Bigg\{\gamma^{\ell}\Bigg[\frac{\vert E_{p,+1}\vert^{2}}{4}(\gamma^{\mu}+\gamma^{\mu}\gamma^{0}+\gamma^{0}\gamma^{\mu}+\gamma^{0}\gamma^{\mu}\gamma^{0}) \nn \\
&\quad+\frac{E_{p,+1}E_{p,-1}^{\ast}}{4}(\gamma^{\mu}-\gamma^{\mu}\gamma^{0}+\gamma^{0}\gamma^{\mu}-\gamma^{0}\gamma^{\mu}\gamma^{0}) \nn \\
&\quad+\frac{E_{p,-1}E_{p,+1}^{\ast}}{4}(\gamma^{\mu}-\gamma^{\mu}\gamma^{0}-\gamma^{0}\gamma^{\mu}-\gamma^{0}\gamma^{\mu}\gamma^{0}) \nn \\
&\quad\frac{\vert E_{p,-1}\vert^{2}}{4}(\gamma^{\mu}+\gamma^{\mu}\gamma^{0}-\gamma^{0}\gamma^{\mu}+\gamma^{0}\gamma^{\mu}\gamma^{0})\Bigg]\Bigg\} \nn \\
&\quad+m\,{\rm Tr}\left\{\left(\frac{\vert E_{p,+1}\vert^{2}}{2}+\frac{\vert E_{p,-1}\vert^{2}}{2}\right)\gamma^{\ell}+\left(\frac{\vert E_{p,+1}\vert^{2}}{2}-\frac{\vert E_{p,-1}\vert^{2}}{2}\right)\gamma^{\ell}\gamma^{0}\right\}\Bigg].
\end{align}
Then, performing the traces with the aid of the identities
\be
{\rm Tr}\{\gamma^{\ell}\gamma^{\mu}\}=-2\delta^{\ell\mu},\qquad  {\rm Tr}\{\gamma^{\ell}\gamma^{0}\gamma^{\mu}\gamma^{0}\}={\rm Tr}\{\gamma^{\ell}(2g^{\mu0}-\gamma^{\mu}\gamma^{0})\gamma^{0}\}=2\delta^{\ell\mu}, 
\ee
we have that
\begin{align}
&{\rm Tr}\{\gamma^{\ell}S(z,z')\}= \nn \\
&\int\frac{{\rm d}^{3}p}{p^{2}-m^2}\Bar{p}_{\mu}\left[\frac{\vert E_{p,+1}\vert^{2}}{4}(-2\delta^{\ell\mu}+2\delta^{\ell\mu})+\frac{E_{p,+1}E_{p,-1}^{\ast}}{4}(-2\delta^{\ell\mu}-4i\varepsilon^{0\mu\ell}-2\delta^{\ell\mu})\right. \nn \\
&\qquad\left.+\frac{E_{p,-1}E_{p,+1}^{\ast}}{4}(-2\delta^{\ell\mu}-4i\varepsilon^{0\mu\ell}-2\delta^{\ell\mu})+\frac{\vert E_{p,-1}\vert^{2}}{4}(-2\delta^{\ell\mu}+2\delta^{\ell\mu})\right].
\end{align}

Now, by taking $\Bar{p}^\mu=(p_{0},0,\sqrt{k})$, it follows that
\be
{\rm Tr}\{\gamma^{\ell}S(z,z')\}=\int\frac{{\rm d}^{3}p}{p^{2}-m^2}\alpha^{(\ell)}[E_{p,+1}^{\ast}E_{p,-1}+E_{p,+1}^{\ast}E_{p,-1}],
\ee
where $\alpha^{(\ell)}=i^{\ell}\sqrt{k}$ for $\ell=1,2$. Therefore,
\be
j^{\ell}=-ie\int\frac{{\rm d}^3p}{p^{2}-m^2}\alpha^{(\ell)}[E_{p,+1}^{\ast}E_{p,-1}+E_{p,+1}^{\ast}E_{p,-1}], \quad \ell=1,2. \label{current density}
\ee
In the next subsection we obtain the charge and current densities for the magnetic field profiles obtained in the previous section.

\subsection{Charge and current density for a seed constant magnetic field}
Inserting the explicit solutions of the Eqs. (\ref{eigenvalor campo constante}) and (\ref{eigenvalor campo constante 1}) into the Eq. (\ref{ec. carga}), we obtain that the charge density to first order intertwining is
\begin{align}
    j^{0}&=-ie\int {\rm d}p_{0}{\rm d}p_{2}\sum_{n=0}^{\infty}\frac{m}{p_{0}^2-k_{n}^{(1)}-m^{2}}(\vert E_{p,+1}\vert^{2}+\vert E_{p,-1}\vert^{2})  \nn \\
    &=\pi e\int {\rm d}p_{2}\Bigg[{\rm sgn}(m)\rho_{0}(x,p_{2})+\sum_{n=0}^{\infty}\frac{m}{\sqrt{m^{2}+(n+1/5)\omega}}\rho_{n+1}(x,p_{2})\Bigg] \nn \\
    &=\pi e\int {\rm d}p_{2}\Bigg[{\rm sgn}(m)\vert\mathcal{N}_{0}^{(1)}\vert^{2}\frac{e^{\eta^{2}}}{\left(\,_{1}F_{1}\left[\frac{1}{10},\frac{1}{2},\eta^2\right]\right)^{2}}+\sum_{n=0}^{\infty}\frac{m}{\sqrt{m^{2}+(n+1/5)\omega}} \nn \\
    &\quad\times\left(\frac{\vert\mathcal{N}_{n+1}^{(1)}\vert^{2}}{2(n+1/5)}\left(\frac{2\eta}{5}\frac{\,_{1}F_{1}\left[\frac{11}{10},\frac{3}{2},\eta^2\right]}{\,_{1}F_{1}\left[\frac{1}{10},\frac{1}{2},\eta^2\right]}F_{n,p_{2},+1}-\sqrt{2n}F_{n-1,p_{2},+1}\right)^{2}+F_{n,p_{2},+1}^{2}\right)\Bigg],
\end{align}
where $\mathcal{N}_{n}^{(1)}$ denotes the corresponding normalization constants of the functions $E_{n,p}^{(1)}(x,y,t)$, and we have used the following result
\be
    \int_{-\infty}^{\infty}\frac{{\rm d}p_{0}}{p_{0}^{2}+b}=\frac{\pi}{\sqrt{b}}.
\ee

Similarly, inserting the explicit solutions of the Eqs. (\ref{eigenvalor campo constante}) and (\ref{eigenvalor campo constante 1}) into the Eq. (\ref{current density}), we obtain that the current density to first order intertwining is
\begin{align}
    j^{\ell}&=-ie\,{\rm Tr}\{\gamma^{\ell}S(z,z')\}=-2i^{\ell+1}e\int {\rm d}p_{0}{\rm d}p_{2}\sum_{n=0}^{\infty}\frac{\sqrt{k_{n}^{(1)}}}{p_{0}^{2}-k_{n}^{(1)}-m^{2}}E_{p,+1}^{\ast}E_{p,-1} \nn \\
    &=-2i^{\ell}e\pi\int {\rm d}p_{2}\sum_{n=0}^{\infty}\frac{\sqrt{(n+1/5)\,\omega}}{\sqrt{m^{2}+(n+1/5)\,\omega}}\,j_{n+1}(x,p_{2})  \nn \\
    &=-2i^{\ell}e\pi\int {\rm d}p_{2}\sum_{n=0}^{\infty}\frac{\sqrt{(n+1/5)\,\omega}}{\sqrt{m^{2}+(n+1/5)\,\omega}}\,\mathcal{N}_{n+1}^{(1)} \nn \\
    &\quad\times\frac{F_{n,p_{2},+1}}{\sqrt{2(n+1/5)}}\left(\frac{2\eta}{5}\frac{\,_{1}F_{1}\left[\frac{11}{10},\frac{3}{2},\eta^2\right]}{\,_{1}F_{1}\left[\frac{1}{10},\frac{1}{2},\eta^2\right]}F_{n,p_{2},+1}-\sqrt{2n}F_{n-1,p_{2},+1}\right).
\end{align}

\begin{figure}[ht]
	\centering
		\includegraphics[width=0.6\textwidth]{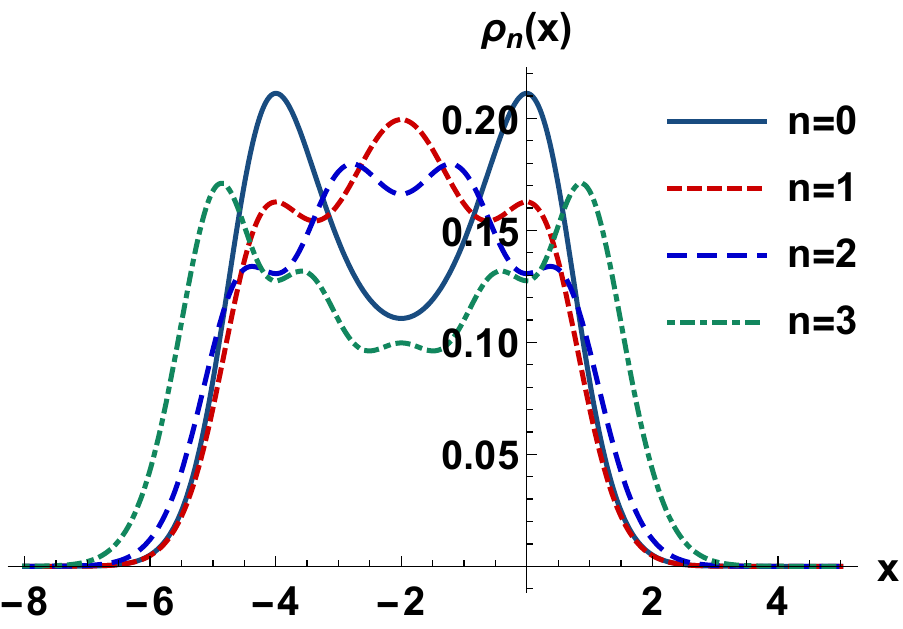}
	\caption{\label{fig:fig3a} Probability density $\rho_{0}(x)$ for the ground state $n=0$ (dark blue, \full) and the excited states $\rho_{n+1}(x)$: $n=0$ (red, \dashed), $n=1$ (blue, \dotted) and $n=2$ (green, \chain). In all the cases $B_{0}=\frac12$, $p_{2}=1$, $\epsilon_{1}=-\frac{\omega}{5}$ and $\omega=1$.}
\end{figure}

\begin{figure}[ht]
	\centering
		\includegraphics[width=0.6\textwidth]{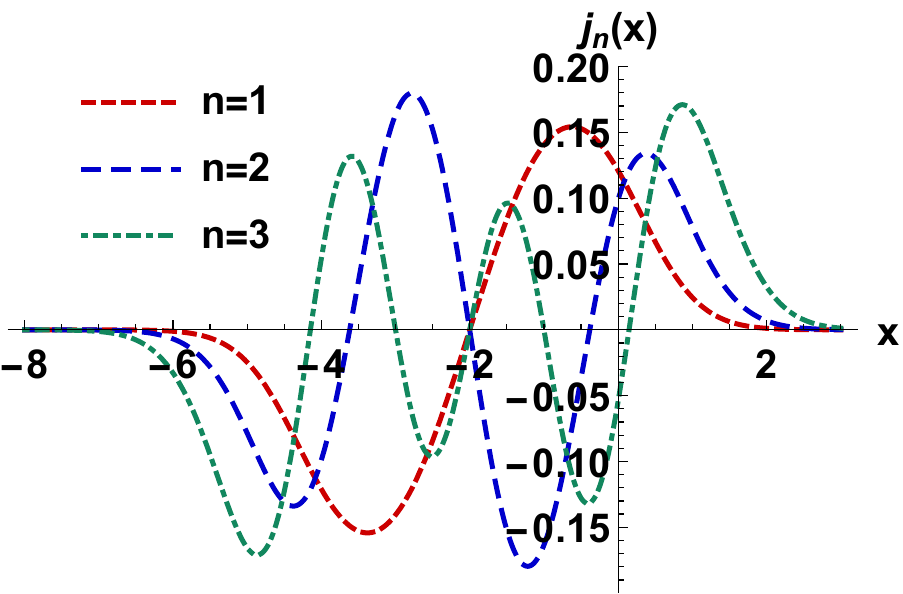}
	\caption{\label{fig:fig3b} Current densities $j_{n+1}(x)$ for the excited states: $n=0$ (red, \dashed), $n=1$ (blue, \dotted) and $n=2$ (green, \chain). In all the cases $B_{0}=\frac12$, $p_{2}=1$, $\epsilon_{1}=-\frac{\omega}{5}$ and $\omega=1$.}
\end{figure}

With these expressions, it is customary to calculate the probability density and probability current. The probability density for the excited states of electrons in graphene is $\rho_{n+1}(x)$ and $\rho_{0}(x)$ for the ground state. The probability currents are $j_{n+1}(x)$ and $j_0(x)=0$, respectively \cite{knn09}. Some graphs for the probability density and probability current are shown in Figs.~\ref{fig:fig3a} and~\ref{fig:fig3b} that agree with \cite{Celeita2020}.

\subsection{Charge and current density for a seed exponentially decaying magnetic field}
Inserting the explicit solutions of the Eqs. (\ref{eigenvalor exponencial}) and (\ref{eigenvalor exponencial 1}) into the Eq. (\ref{ec. carga}), we obtain that the charge density to first order intertwining is
\begin{align}
    &j^{0}=-ie\int {\rm d}p_{0}{\rm d}p_{2}\sum_{n=0}^{\infty}\frac{m}{p_{0}^2-k_{n}^{(1)}-m^{2}}(\vert E_{p,+1}\vert^{2}+\vert E_{p,-1}\vert^{2}) \nn \\
    &=\pi e\int {\rm d}p_{2}\Bigg[{\rm sgn}(m)\rho_{0}(x,p_{2})+\sum_{n=0}^{\infty}\frac{m}{\sqrt{m^{2}+k_{n+1}^{(1)}}}\rho_{n+1}(x,p_{2})\Bigg] \nn \\
    &=\pi e\int {\rm d}p_{2}\Bigg[{\rm sgn}(m)\vert\mathcal{N}_{0}^{(1)}\, F_{0,p_{2}}^{(1)}\vert^{2}+\sum_{n=0}^{\infty}\frac{m}{\sqrt{m^{2}+k_{n+1}^{(1)}}}\left(\vert\mathcal{N}_{n+1}^{(1)}\,F_{n+1,p_{2}}^{(1)}\vert^{2}+\vert F_{n,p_{2},+1}\vert^{2}\right)\Bigg]. \label{carga caso exponencial}
\end{align}

Once again, inserting the explicit solutions of the Eqs. (\ref{eigenvalor exponencial}) and (\ref{eigenvalor exponencial 1}) into the Eq. (\ref{current density}), we obtain that the charge density to first order intertwining is
\begin{align}
    j^{\ell}&=-ie\,{\rm Tr}\{\gamma^{\ell}S(z,z')\}=-2i^{\ell+1}e\int {\rm d}p_{0}{\rm d}p_{2}\sum_{n=0}^{\infty}\frac{\sqrt{k_{n}^{(1)}}}{p_{0}^{2}-k_{n}^{(1)}-m^{2}}E_{p,+1}^{\ast}E_{p,-1} \nn \\
    &=-2i^{\ell}e\pi\int {\rm d}p_{2}\sum_{n=0}^{\infty}\frac{\sqrt{k_{n+1}^{(1)}}}{\sqrt{m^{2}+k_{n+1}^{(1)}}}\,j_{n+1}(x,p_{2}) \nn \\
    &=-2i^{\ell}e\pi\int {\rm d}p_{2}\sum_{n=0}^{\infty}\frac{\sqrt{k_{n+1}^{(1)}}}{\sqrt{m^{2}+k_{n+1}^{(1)}}}\,\mathcal{N}_{n+1}^{(1)}\,F_{n,p_{2},+1}(\rho)\,F_{n+1,p_{2}}^{(1)}(\rho). \label{corriente caso exponencial}
\end{align}

\begin{figure}[ht]
	\centering
		\includegraphics[width=0.6\textwidth]{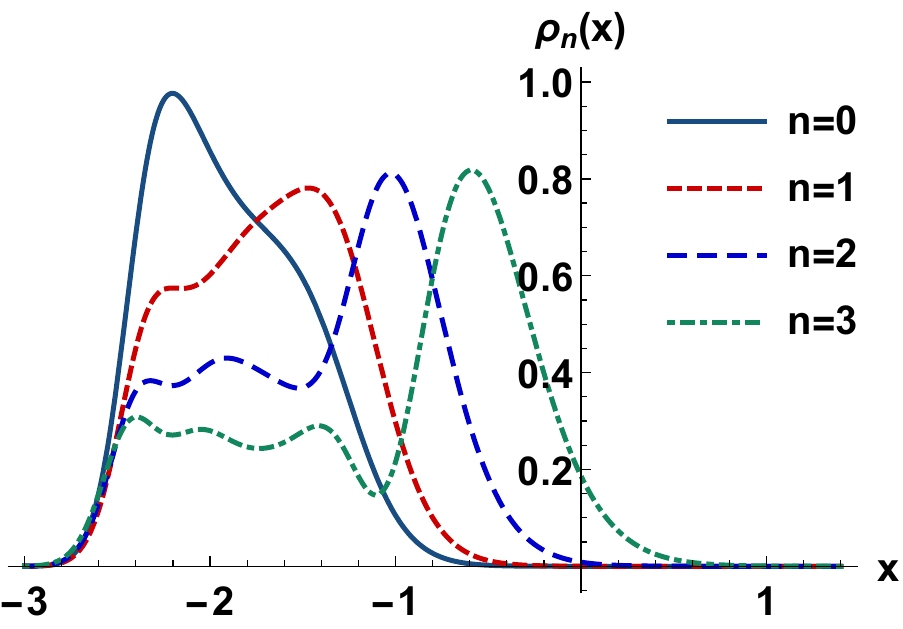}
	\caption{\label{fig:fig4a} Probability density $\rho_{0}(x)$ for the ground state $n=0$ (dark blue, \full) and the excited states $\rho_{n+1}(x)$: $n=0$ (red, \dashed), $n=1$ (blue, \dotted) and $n=2$ (green, \chain). In all the cases $B_{0}=1$, $\nu_1=-\frac{3}{2}$, $p_{2}=5\alpha$, $\epsilon_{1}=-\frac{11\alpha^{2}}{2}$ and $\alpha=1$.}
\end{figure}

\begin{figure}[ht]
	\centering
		\includegraphics[width=0.6\textwidth]{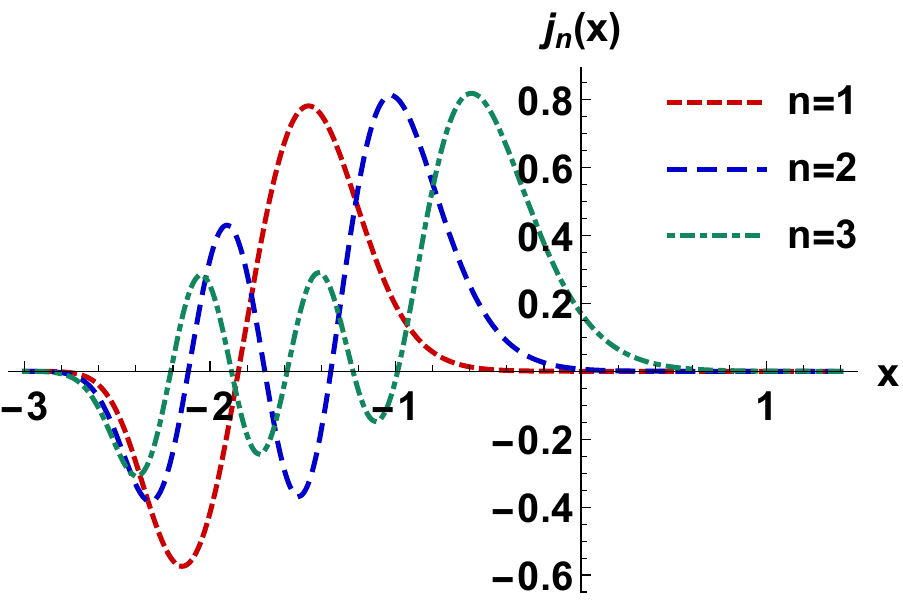}
	\caption{\label{fig:fig4b} Current densities $j_{n+1}(x)$ for the excited states: $n=0$ (red, \dashed), $n=1$ (blue, \dotted) and $n=2$ (green, \chain). In all the cases $B_{0}=1$, $\nu_1=-\frac{3}{2}$, $p_{2}=5\alpha$, $\epsilon_{1}=-\frac{11\alpha^{2}}{2}$ and $\alpha=1$.}
\end{figure}

Thus, we calculate the probability density and probability current with these expressions, which are plotted in Figs.~\ref{fig:fig4a} and~\ref{fig:fig4b}.

\subsubsection{Behavior for small inhomogeneity $\alpha$}
Let us consider the asymptotic behavior of the previous results for small values of the parameter $\alpha$ in order to compare them with those for the constant magnetic field case, since in the limit $\alpha\rightarrow0$ the exponentially decaying magnetic field tends to the constant magnetic field. 

\begin{figure}[ht]
	\centering
	\begin{minipage}[b]{0.48\textwidth}
		\includegraphics[width=\textwidth]{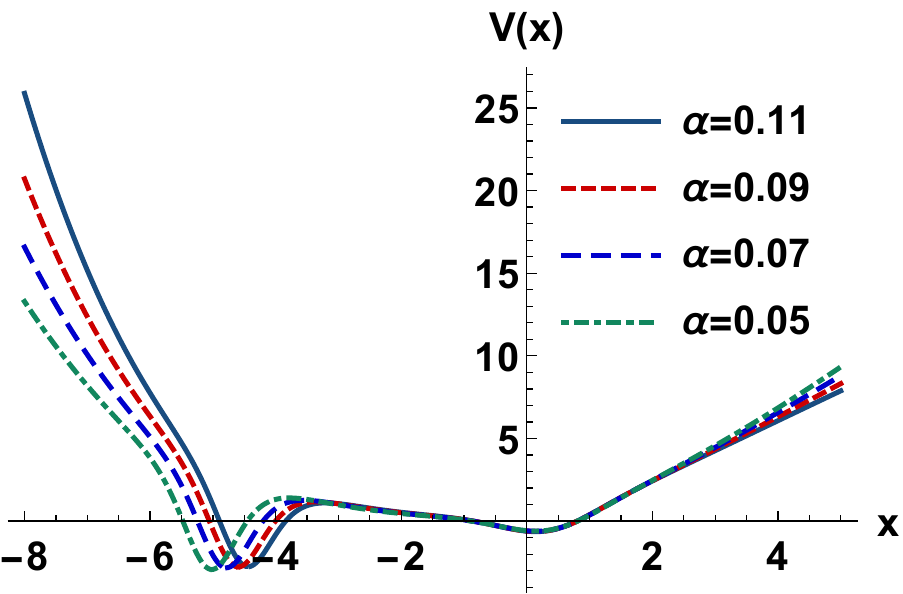}\\
		\centering{\footnotesize (a)}
		\label{fig:fig5a}
	\end{minipage}
	\hspace{0.25cm}
	\begin{minipage}[b]{0.48\textwidth}
		\includegraphics[width=\textwidth]{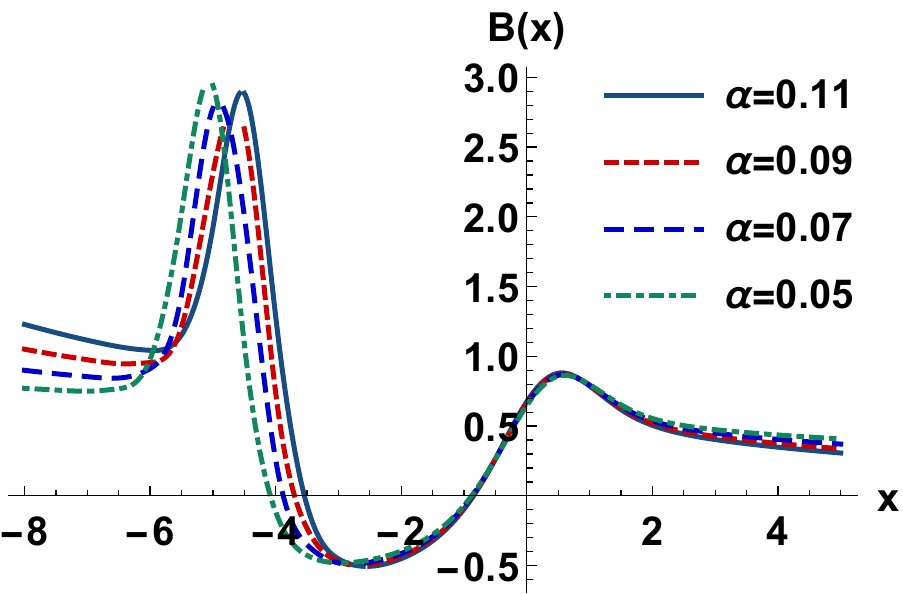}\\
		\centering{\footnotesize (b)}
		\label{fig:fig5b}
	\end{minipage}
	\caption{\label{fig:fig5}(a) Generated potential $V_1(x,\epsilon_1)$ in~(\ref{SUSYpotexp}) and (b) generated magnetic field $B_1(x, \epsilon_1)$ in~(\ref{SUSYmagnexp}) for small inhomogeneity ($\alpha<1$): $\alpha=0.11$ (dark blue, \full), $\alpha=0.09$ (red, \dashed), $\alpha=0.07$ (blue, \dotted) and $\alpha=0.05$ (green, \chain). In all the cases $B_{0}=\frac12$, $\nu_1=-\frac{3}{2}$, $p_{2}=1$ and $\epsilon_{1}=-\frac{k_{1}^{+}}{5}$.}
\end{figure}

\begin{figure}[ht]
	\centering
		\includegraphics[width=0.6\textwidth]{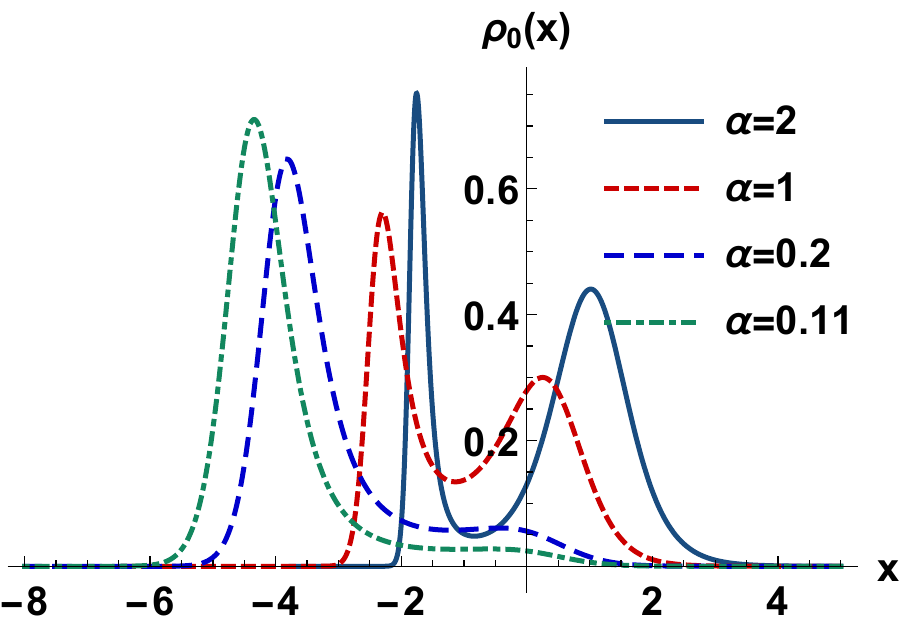}\\
	\caption{\label{fig:fig8}Probability density $\rho_0(x)$ for the ground state $n=0$ in~(\ref{eigenvalor exponencial}) for different values of inhomogeneity: $\alpha=2$ (dark blue, \full), $\alpha=1$ (red, \dashed), $\alpha=0.2$ (blue, \dotted) and $\alpha=0.11$ (green, \chain). In all the cases $B_{0}=\frac12$, $\nu_1=-\frac{3}{2}$, $p_{2}=1$ and $\epsilon_{1}=-\frac{k_{1}^{+}}{5}$.}
\end{figure}

\begin{figure}[htbp]
	\centering
	\begin{minipage}[b]{0.38\textwidth}
		\includegraphics[width=\textwidth]{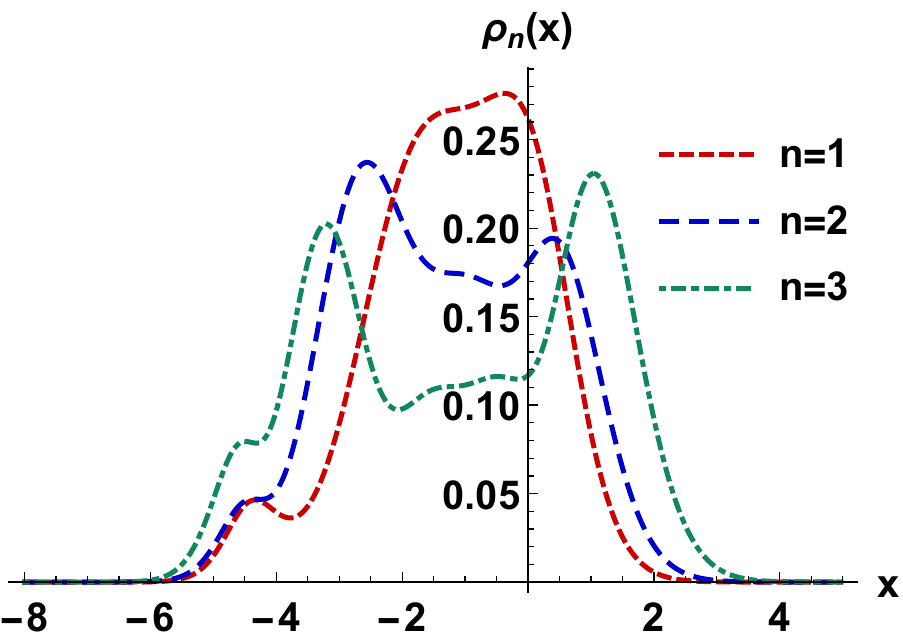}\\
		\centering{\footnotesize (a) $\alpha=0.11$}
		\label{fig:fig7a}
	\end{minipage}
	\hspace{1.5cm}
	\begin{minipage}[b]{0.38\textwidth}
		\includegraphics[width=\textwidth]{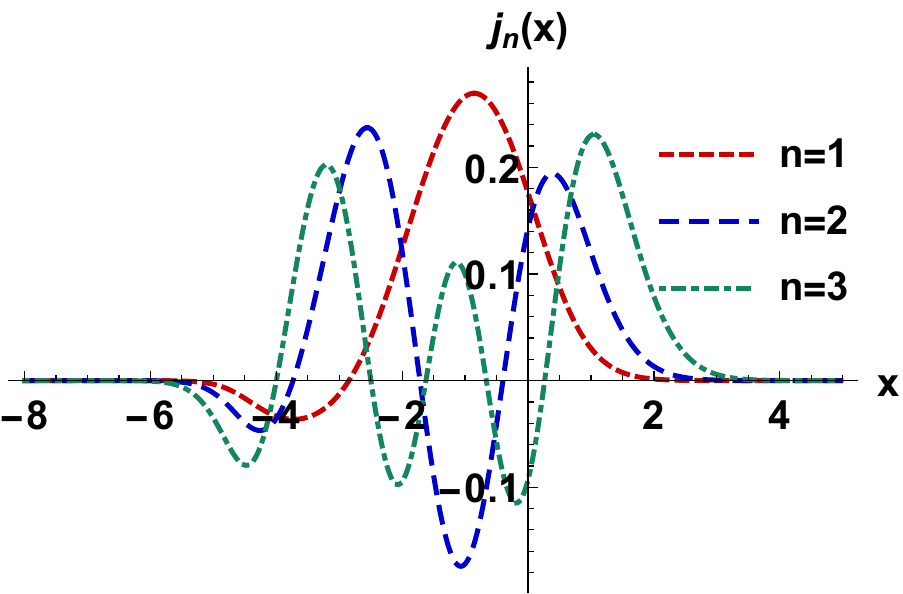}\\
		\centering{\footnotesize (b) $\alpha=0.11$}
		\label{fig:fig7b}
	\end{minipage}
	\begin{minipage}[b]{0.38\textwidth}
		\includegraphics[width=\textwidth]{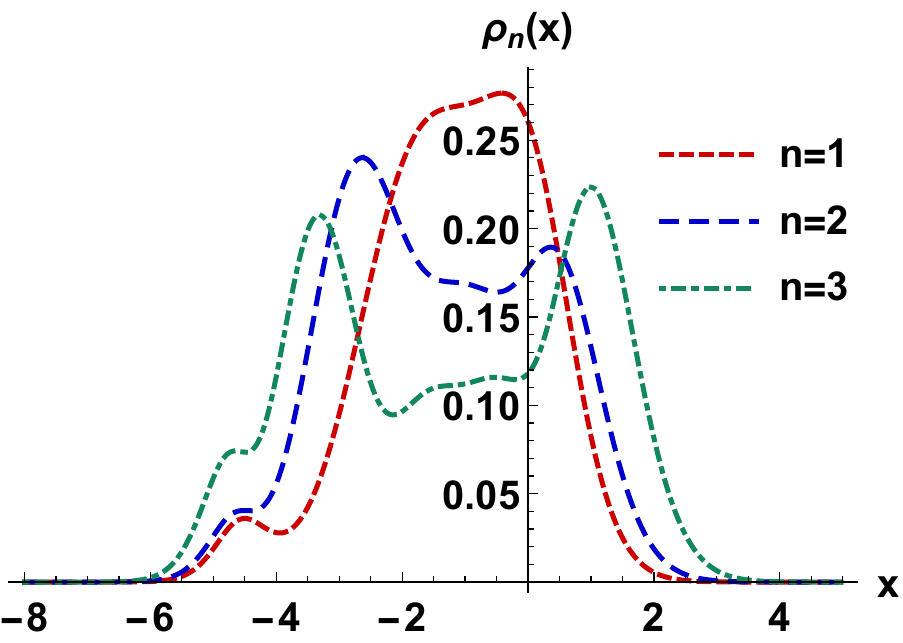}\\
		\centering{\footnotesize (c) $\alpha=0.09$}
		\label{fig:fig7c}
	\end{minipage}
	\hspace{1.5cm}
	\begin{minipage}[b]{0.38\textwidth}
		\includegraphics[width=\textwidth]{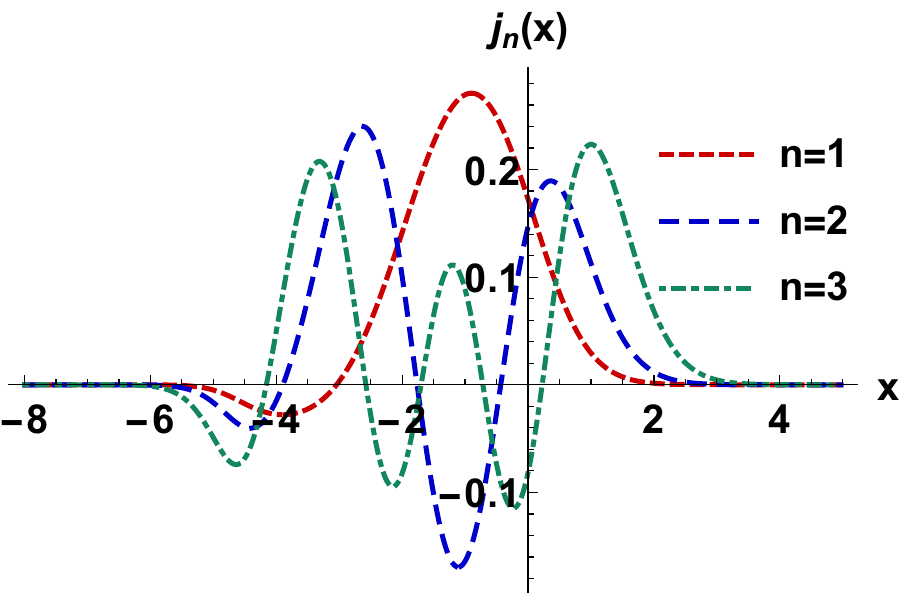}\\
		\centering{\footnotesize (d) $\alpha=0.09$}
		\label{fig:fig7d}
	\end{minipage}
	\begin{minipage}[b]{0.38\textwidth}
		\includegraphics[width=\textwidth]{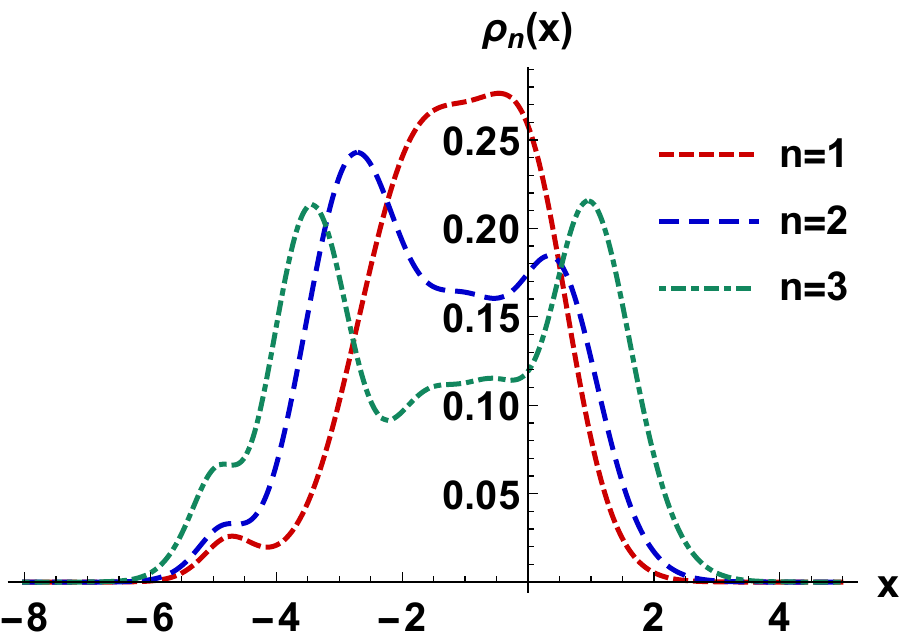}\\
		\centering{\footnotesize (e) $\alpha=0.07$}
		\label{fig:fig7e}
	\end{minipage}
	\hspace{1.5cm}
	\begin{minipage}[b]{0.38\textwidth}
		\includegraphics[width=\textwidth]{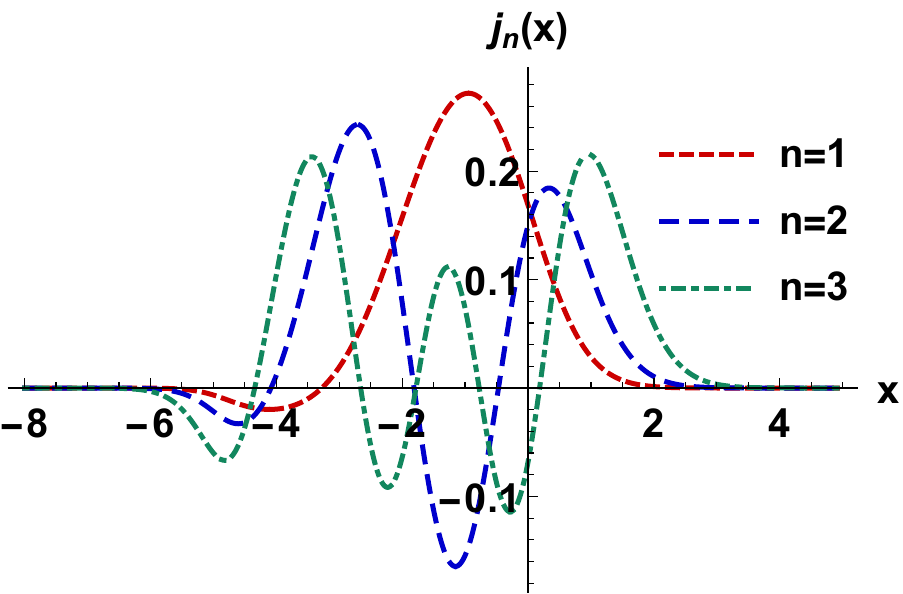}\\
		\centering{\footnotesize (f) $\alpha=0.07$}
		\label{fig:fig7f}
	\end{minipage}
\begin{minipage}[b]{0.38\textwidth}
		\includegraphics[width=\textwidth]{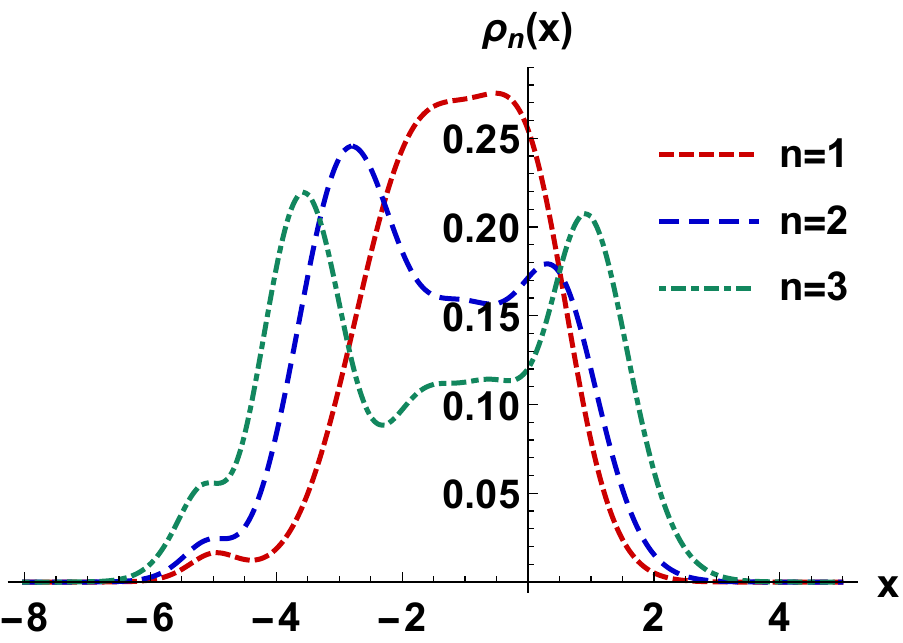}\\
		\centering{\footnotesize (g) $\alpha=0.05$}
		\label{fig:fig7g}
	\end{minipage}
	\hspace{1.5cm}
	\begin{minipage}[b]{0.38\textwidth}
		\includegraphics[width=\textwidth]{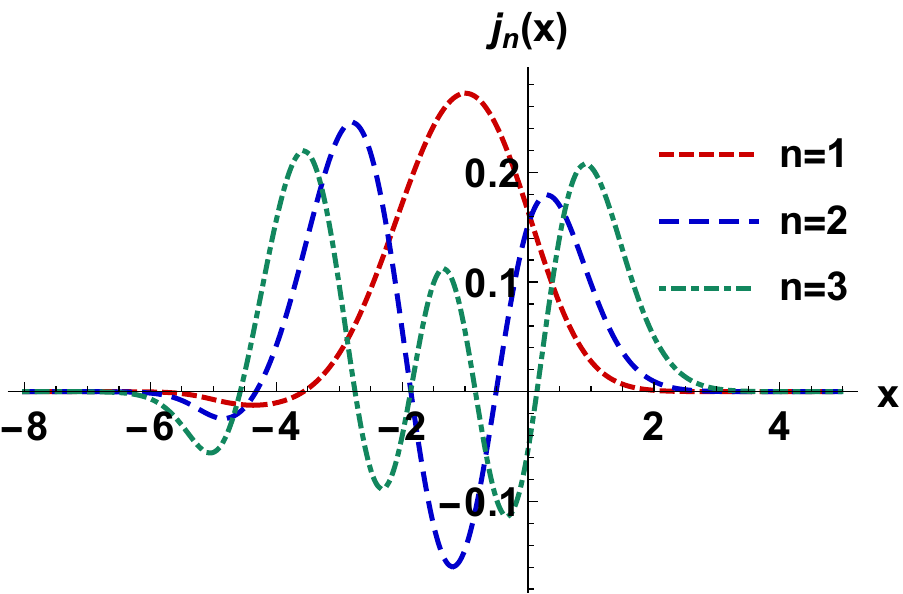}\\
		\centering{\footnotesize (h) $\alpha=0.05$}
		\label{fig:fig7h}
	\end{minipage}
	\caption{\label{fig:fig7}Probability density $\rho_0(x)$ (left-hand) and current densities $j_{n+1}(x)$ (right-hand) for the excited states in~(\ref{eigenvalor exponencial 1}): $n=0$ (red, \dashed), $n=1$ (blue, \dotted) and $n=2$ (green, \chain). In all the cases $B_{0}=\frac12$, $\nu_1=-\frac{3}{2}$, $p_{2}=1$, $\epsilon_{1}=-\frac{k_{1}^{+}}{5}$ and small inhomogeneity ($\alpha<1$).}
\end{figure}

Indeed, rewriting $D=\omega/(2\alpha)$, being $\omega$ as in Eq.~(\ref{Wpotential}), the superpotential $W_{0}(x)$, the Morse potential $V_{0}^{-}(x)$ and the eigenenergies $k_{n}^{+}$ in Eqs.~(\ref{superpotentialexp}),~(\ref{Morsepotential}) and~(\ref{spectrumexp}), respectively, turn into:
\begin{subequations}
\begin{align}
\lim_{\alpha\rightarrow0} W_{0}(x) &= \lim\limits_{\alpha\rightarrow0} p_{2}-\frac{\omega}{2\alpha}(\exp\left(-\alpha x\right)-1)=\frac{\omega}{2}x+p_{2}, \label{eq. W0} \\
\lim_{\alpha\rightarrow0} V_{0}^{-}(x) &= \lim\limits_{\alpha\rightarrow0} \left(p_{2}+\frac{\omega}{2\alpha}\right)^{2}+\frac{\omega^{2}}{4\alpha^{2}}\exp\left(-2\alpha x\right)-\frac{\omega}{\alpha}\left(p_{2}+\frac{\omega}{2\alpha}+\frac{\alpha}{2}\right)\exp\left(-\alpha x\right) \nonumber \label{eq. V0-}\\
&= \frac{\omega^2}{4}\left(x+\frac{2p_{2}}{\omega}\right)^{2}-\frac{\omega}{2}, \\
\lim_{\alpha\rightarrow0}k_{n}^{+} &= \lim\limits_{\alpha\rightarrow0}\alpha\, n\left(2\left[p_{2}+\frac{\omega}{2\alpha}\right]-\alpha\,n\right)=\omega\, n, \quad n=0,1,\dots \label{eq. kn+}
\end{align}
\end{subequations}
which coincide with Eqs.~(\ref{Wpotential}),~(\ref{HOpotential}) and~(\ref{spectrumconst}), respectively. In Fig.~\ref{fig:fig5}, we show the behavior of the potential $V_{1}(x,\epsilon_{1})$ and the magnetic field profile $B_{1}(x,\epsilon_{1})$ for small values of $\alpha$. According to the plots, for small inhomogeneity the potential and the magnetic field generated by the supersymmetric transformation have a similar behavior to their counterparts in the constant magnetic field case for $\vert x\vert\rightarrow\infty$, while for values of $x$ in the inner region of $V_{1}(x,\epsilon_{1})$ and $B_{1}(x,\epsilon_{1})$, such functions do not look like those in Fig.~\ref{fig:fig1}.

On the other hand, the probability density $\rho_{0}(x)$ does not look like the corresponding one for the constant magnetic field case, as shown in Fig.~\ref{fig:fig8}. This suggests us that after the supersymmetric transformation is implemented to the exponentially decaying magnetic field case, we are not able to recover the eigenfunction $F_{0,p_{2}}^{(1)}\sim 1/u_{1}$ for the ground state of the Hamiltonian $\mathcal{H}_{1}$ of the constant magnetic field case. Likewise, the corresponding probability density and probability current for small values of $\alpha$ and $\epsilon_{1}=-k_{1}^{+}/5$ are shown in Fig.~\ref{fig:fig7}. As we can see, the probability densities $\rho_{n}(x)$ and the current densities $j_{n}(x)$ for the excited states with $n>1$ exhibit a subtle resemblance to the plots shown in Figs.~\ref{fig:fig3a} and~\ref{fig:fig3b} for a constant magnetic field $B_{0}=1/2$ and $p_{2}=1$. However, such functions are asymmetric respect to $x=-2$ in comparison with those that correspond to the eigenfunctions in Eqs.~(\ref{eigenvalor campo constante}) and~(\ref{eigenvalor campo constante 1}).

\section{Final remarks}\label{sec5}
In this work, we have studied the Dirac fermion propagator for graphene-like systems in external magnetic fields. We have constructed the Dirac fermion propagator for graphene-like systems in the presence of non-trivial and inhomogeneous external magnetic fields generated by first-order intertwining operators from the seed solutions corresponding to the Ritus eigenfunctions for a constant magnetic field and an exponetially decaying magnetic field~\cite{Concha,knn09, Milpas,Jakubsky,Jakubsky1,Jahani} already known in literature. We constructed the propagator in  the basis of the eigenfunctions of the operator $(\gamma \cdot \Pi)^2$ for new non-trivial magnetic field profiles, hence extending the number of cases in which the propagator admits a closed form representation.

The generalized first-order intertwining method presented here have been followed of the discussion in Ref.~\cite{Midya2014,Celeita2020}. By choosing the parameters $\epsilon_{1}=-k_{1}^{+}/5=-\omega/5$ and $\nu_{1}=0$, we have obtained the generated potential $V_{1}(x, \epsilon_{1})$ and the magnetic field profile $B_{1}(x,\epsilon_{1})$ for the case of the seed uniform magnetic field whose graphs have been plotted in Fig. \ref{fig:fig1} and that agree with \cite{Celeita2020}. Similarly, taking $\epsilon_{1}=-k_{1}^{+}/2=-\alpha(2q_{2}-\alpha)/2$ and the restriction $\nu_{1}\in\mathbb{R}-\{[-1,0]\}$, we obtained the generated potential $V_{1}(x, \epsilon_{1})$ and the magnetic field profile $B_{1}(x,\epsilon_{1})$ for the case of the seed exponentially decaying magnetic field whose graphs have been plotted in Fig. \ref{fig:fig2} and that agree with \cite{Celeita2020} too. In both cases, the energy spectrum of the corresponding Hamiltonian $\mathcal{H}_{1}$ has one more level than $\tilde{\mathcal{H}}_{0}$ derived from the selection of the parameter $\epsilon_{1}$ and the function $u_{1}$.

On the other hand, we have found the charge and current densities from the constructed Dirac fermion propagator in the non-trivial  examples of inhomogeneous fields derived from the intertwining framework. From such densities, the probability density and the probability current for both cases of fields have been obtained and plotted in Figs.~\ref{fig:fig3a}, \ref{fig:fig3b}, \ref{fig:fig4a} and \ref{fig:fig4b} for the ground state $n=0$ and some excited states that reproduce the densities in the literature \cite{Celeita2020} from the direct solutions of the wave equation in the said background fields. About the number of nodes observed in the plots of the probability densities $\rho_{n+1}(x)$ and current densities $j_{n+1}(x)$, and its relation with index $n$, let us point out that the wave functions $F_{n,p_2,+1}$ and $F_{n,p_2,-1}$ satisfy, individually, the node theorem but the probability density $j^0$ not necessary does it, since it depends how such functions overlap. The current density $j^\ell$ also mixes functions $F_{n,p_2,+1}$ and $F_{n,p_2,-1}$, so that node pattern in such a current can be traced back to the individual behavior of these functions, but the product and sum of individual contributions might not follow an obvious node pattern. Something similar occurs when the 1-SUSY QM formalism is applied to generate the functions $F_{n,p_2}^{(1)}$, which no necessary satisfy the node theorem.

Additionally, we have shown that for the limit $\alpha\rightarrow0$ the exponentially decaying magnetic field tends to the constant magnetic field. Thus, taking the limit $\alpha\rightarrow0$ in the exponentially decaying magnetic field Morse potentials (Eqs.~(\ref{superpotentialexp}),~(\ref{Morsepotential})) and eigenenergies (Eq.~(\ref{spectrumexp})), respectively, we have obtained the Morse potential $V_{0}^{+}(x)$ (Eqs.~(\ref{eq. V0-}), (\ref{eq. W0})) and the eigenenergies  $k_{n}^{+}$ (Eq. (\ref{eq. kn+})) that coincide with the superpotential (Eqs. (\ref{Wpotential}),~(\ref{HOpotential})) and eigenenergies (Eq. (\ref{spectrumconst})) of the uniform magnetic field case, respectively. Also, we have shown that for small inhomogeneity $\alpha$ the behavior of the potential $V_{1}(x,\epsilon_{1})$ and the magnetic field profile $B_{1}(x,\epsilon_{1})$ in general does not coincide with the uniform magnetic field case, except in the asymptotic limit $\vert x\vert\rightarrow\infty$.    

Moreover, we have plotted the probability density taking the limit $\alpha\rightarrow0$ in order to recover the probability density of the uniform magnetic field case, however this does not occur with our results as the Fig.~\ref{fig:fig8} shows because after the supersymmetric transformation is implemented to the exponentially decaying magnetic field case, we have not recovered the eigenfuntion $F_{0,p_{2}}^{(1)}\sim 1/u_{1}$ for the ground state of the Hamiltonian $\mathcal{H}_{1}$  of the constant magnetic field case. For small values of $\alpha$ and $\epsilon_{1}=-k_{1}^{+}/5$ we have obtained the probability density and probability current as shown in Fig.~\ref{fig:fig7} where the plots are not symmetric respect to $x=-2$, as occurs in Figs.~\ref{fig:fig3a} and~\ref{fig:fig3b} for the excited states.

For the future, we are planning  to obtain the Ritus functions for graphene-like systems for the fields studied in this work by second-order intertwining operators. Results will be reported elsewhere. This work is expected to  become a guide for colleagues interested in the theoretical developments of graphene-like systems.  

\section*{Acknowledgments}
EDB and AR acknowledge financial support from CONACYT Project FORDECYT-PRONACES/61533/2020. EDB also acknowledges the SIP-IPN research grant 20220025. YCS acknowledges the CIC-UMSNH research grant 6297771/2021.

\bibliographystyle{ieeetr}
\bibliography{biblio}

\end{document}